\documentclass[12pt,preprint]{aastex}

\slugcomment{Accepted for publication in ApJ}

\shorttitle{Variability of the Accreting Brown Dwarf 2M1207}
\shortauthors{Stelzer et al.}

\begin{document}


\title{Emission Line Variability of the Accreting Young Brown Dwarf 2MASSW J1207334-393254: From Hours to Years}


\author{Beate Stelzer}
\affil{INAF -- Osservatorio Astronomico di Palermo, Piazza del Parlamento 1, 90134 Palermo, Italy}
\email{stelzer@astropa.unipa.it}

\and 

\author{Alexander Scholz}
\affil{SUPA, School of Physics and Astronomy, University of St.Andrews, North Haugh, St. Andrews, Fife KY\,16~9SS, United Kingdom}

\and 

\author{Ray Jayawardhana}
\affil{Department of Astronomy \& Astrophysics, University of Toronto, 50 St. George Street, Toronto, ON\,M5S\,3H4, Canada}



\begin{abstract}
We have obtained a series of high-resolution optical spectra for 2MASSW J1207334-393254 
(2M\,1207).
Two consecutive observing nights at 
the ESO Very Large Telescope with the UVES spectrograph yielded a timeseries with a 
resolution of $\sim 12$\,min.
Additional high-resolution optical spectra were obtained months 
apart at the Magellan Clay telescope using the MIKE instrument. Combined with previously 
published results, these data allow us to investigate changes in the emission line 
spectrum of 2M\,1207 on timescales of hours 
to years. Most of the emission line profiles of 2M\,1207 are broad, 
in particular that of H$\alpha$, indicating that the dominant fraction of the emission 
must be attributed to disk accretion rather than to magnetic activity. From the 
H$\alpha$ $10$\,\% width we deduce a relatively stable accretion rate between 
$10^{-10.1...-9.8}\,{\rm M_\odot/yr}$ for two nights of consecutive observations.
Therefore, either the accretion stream is nearly homogeneous over (sub-)stellar longitude  
or the system is seen face-on. 
Small but significant variations are evident throughout our 
near-continuous observation, and they reach a maximum after $\sim 8$\,h, roughly the 
timescale on which maximum variability is expected across the rotation cycle. Together 
with past measurements, we confirm that the accretion rate of 2M\,1207 varies by more 
than one order of magnitude on timescales of months to years. Such variable mass accretion 
yields a plausible explanation for the observed spread in the $\dot{M} - M$ diagram. 
The magnetic field required to drive the funnel flow is on the order of a few 
hundred G. 
Despite the obvious presence of a magnetic field, no radio nor X-ray emission 
has been reported for 2M\,1207. Possibly strong accretion suppresses magnetic 
activity in brown dwarfs, similar to the findings for higher mass T Tauri stars. 
\end{abstract}


\section{Introduction}

In recent years, there is increasing evidence that young brown dwarfs evolve similarly to 
their higher-mass counterparts, the T Tauri stars (TTS). 
In particular, their formation process includes
a phase of mass accretion from a circumstellar disk, 
whose presence is probed by infrared emission in excess of the photosphere 
\citep[e.g., ][]{Natta02.1, Jayawardhana03.1, Luhman05.3, Allers06.1}. 
Spectroscopic features, such as the strength and shape of H$\alpha$ emission, directly
trace the accreting material. 
In the stellar regime, `weak TTS' have no optically thick circumstellar matter, 
and their H$\alpha$ emission is purely chromospheric. 
In `classical TTS', the line emission is dominated by contributions from accretion. Next to
displaying much stronger lines, their line profiles are broader due to the high velocities in
the accreting gas and Stark broadening \citep{Muzerolle01.1}. Winds may play a role as well,
and are recognized by blueshifted absorption components and forbidden emission lines
\citep{Mundt84.1, Edwards87.1, Hamann94.1, Hartigan95.1}.
Analogous observations have been made for brown dwarfs, where classical TTS-like H$\alpha$ emission
persists down to the deuterium burning limit \citep{Mohanty05.1}. Forbidden emission lines, 
tracing outflows have been reported from a small number of brown dwarfs 
\citep[e.g.,][]{Fernandez01.1,Barrado04.4}. 

Brown dwarfs are intrinsically faint objects, therefore a detailed study of magnetic 
activity and accretion in young brown dwarfs is restricted to the most nearby ones. 
A particularly well-suited playground is the TW Hya association (TWA), with
four confirmed brown dwarfs at an age of $\sim 8$\,Myr and a distance of $\sim 50$\,pc. 
%
2M\,1207 is the best-studied among the brown dwarfs in the TWA. 
Its youth was confirmed by low gravity 
and Lithium absorption, 
and its TWA membership by radial velocity 
and proper motion 
studies \citep{Gizis02.2, Mohanty03.2, Scholz05.5}. 
\cite{Mamajek05.1} has revised the previous distance estimate of 2M\,1207 using 
the moving cluster method, resulting in $53 \pm 6$\,pc. 
2M\,1207 is the oldest brown dwarf known to 
actively accrete from a disk, thus representing a benchmark in constraining 
the disk lifetimes in the substellar regime. 
The object has recently aroused strong interest because of the discovery 
of a planetary mass companion \citep{Chauvin04.1, Chauvin05.1}. 
This companion is at a separation of $40$\,AU and not 
expected to influence the magnetic and accretion activity of the brown dwarf.

Stellar astronomers are intrigued by 2M\,1207 being the closest known brown dwarf with a disk. 
No substantial $L^{\prime}$ excess was detected by \cite{Jayawardhana03.1}, 
but mid-IR photometry established the presence of circumstellar material \citep{Sterzik04.1}.
This is also corroborated by {\em Spitzer} photometry, where 2M\,1207 is detected 
in all IRAC bands and at $24\,\mu$m above photospheric levels \citep{Riaz06.1}.
The spectral energy distribution of 2M\,1207 shows evidence for dust evolution,
similar to the higher-mass disk-bearing TTS in the TWA. 
2M\,1207 is therefore a lower-mass analog of the classical TTS. 

Strong H$\alpha$ emission has been reported right with the identification of 2M\,1207
as a TWA candidate \citep{Gizis02.2}. 
Subsequent
high-resolution spectra have shown that the H$\alpha$ profile is broadened and
asymmetric, thus displaying characteristic signatures of accretion \citep{Mohanty03.2}. 
The H$\alpha$ equivalent width measured by \cite{Mohanty03.2} was only
about one-tenth of the earlier value obtained by \cite{Gizis02.2},  
and this variability was interpreted as a sign for variable mass accretion rate. 
In a subsequent study of emission line variability, aiming at constraining these variations, 
2M\,1207 presented dramatic changes in the shape and intensity of its H$\alpha$ line in 
data obtained two months apart 
\citep{Scholz05.4}.
The observed profile variability was consistent with a scenario of asymmetric accretion
with a direct view into the (magnetic) funnel flow
\citep{Scholz06.1}.
In this picture, variability is also expected on the time scale of the rotation
period ($\sim 1$\,d),
as the hot accretion spots are corotating with the star. Existing data have remained inconclusive
in this respect due to poor sampling. Therefore, we have collected new high spectral resolution
data with the aim to monitor the H$\alpha$ variability on shorter time scales. 

Here we present the results obtained during two consecutive nights of observation, and
from five additional spectra taken weeks to months apart. 
The observations are described in Sect.~\ref{sect:obs}. 
In Sect.~\ref{sect:profiles} we present a detailed analysis of the characteristics of 
the H$\alpha$ emission and other prominent emission lines in the optical spectrum of 2M\,1207. 
The results are discussed in Sect.~\ref{sect:discussion}, where the spectral variability
is examined on various timescales from hours to years. 
Estimates for the accretion rate and the magnetic field strength of 2M\,1207 are presented, 
and put in context with other (sub)stellar objects. Finally, we compare the accretion and
activity characteristics of all known brown dwarfs in TWA. 
Concluding remarks are found in Sect.~\ref{sect:conclusions}.

\section{Observations}\label{sect:obs}

2M\,1207 was observed with the {\em Ultraviolet and Visual Echelle Spectrograph} (UVES) 
at the ESO VLT\,2 during two nights May 8-9, 2006 in the dichroic mode that covers 
the full spectral range of UVES by simultaneous use of both the blue and the red arm. 
We used the standard setting DICHR\,\#2 centered at $437$\,nm for the
blue arm and at $760$\,nm for the red arm. 
This setup yields simultaneous spectral coverage from  $3730-4990$\,\AA~ on the blue chip 
and from $5650-9460$\,\AA~ on the red mosaic. 
A slit width of $0.8^{\prime\prime}$ was chosen 
resulting in a resolving power $R \approx 50 000$. 
The CCDs were rebinned to $2 \times 2$ pixels.

The integration time was $600$\,s. During the two observing nights, we obtained a 
total of $59$ exposures of 2M\,1207 at blue and red wavelengths. Except for 
some constraints due to calibrations and weather the frames were taken consecutively, 
such that the time resolution for the investigation of spectral changes is $\sim 12$\,min 
over a large part of the observations.


The data were processed with the UVES pipeline implemented in the MIDAS environment 
\citep{Ballester00.1}. 
Reduction steps included bias subtraction, flatfield correction, and wavelength calibration. 
The final one-dimensional spectra have been corrected for contaminating sky light
and cosmic rays have been removed. 
Each spectrum was corrected for the radial velocity of 2M\,1207 
\citep[$11.2$\,km/s; ][]{Mohanty03.2} and for the barycenter motion of the Earth. 

We extracted a spectrum from each of the individual science frames. In addition, 
to increase the S/N for the detection and measurements of weak spectral features, we built
averages of each four consecutive frames. Thus, the effective exposure times of these merged spectra 
are $40$\,min. 
Three individual frames with poor S/N were not considered
in the merging process, such that we obtain $14$ average spectra from the two nights.
The spectra were rebinned to $R=50000$ around individual emission lines 
before the scientific analysis was carried out. 

In February and April 2006, we obtained five additional exposures of 2M\,1207 using 
the Magellan Inamori Kyocera Echelle (MIKE) spectrograph at the Magellan Clay telescope 
on Las Campanas, Chile. These data were reduced using standard routines within IRAF, 
including bias subtraction, flatfielding, wavelength and relative flux calibration,
and removal of cosmic rays. For more details of the reduction procedure, see
\cite{Scholz06.1}. Exposure times for the MIKE spectra were
typically $10$\,min, and the nominal resolution is $R \approx 25000$.

\section{Emission line spectrum}\label{sect:profiles}

\subsection{H$\alpha$ emission}\label{subsect:halpha}

The main aim of this study is to examine the variability of the H$\alpha$ emission of 2M\,1207
with good time resolution and on time-scales of $\sim 1$\,d, corresponding to its 
suspected rotation period. An estimate for the period is given by \citet{Scholz05.4}. In brief, 
the rotational velocity of 2M\,1207 is $v \sin{i} = 13 \pm 2$\,km/s \citep{Mohanty03.2}. 
Together with a radius of $0.27\,R_\odot$ extracted from evolutionary models \citep{Chabrier00.2},
the period is expected to be $P / \sin{i} \approx 25.2$\,h. Assuming a random position for
the inclination, the most probable value for $\sin{i}$ is $0.81$, and for this value the rotation period
is $20.4$\,h. Magnetospheric accretion models \citep{Hartmann94.1} have shown that 
the presence of a red absorption feature in the Balmer lines (frequently seen in 2M\,1207; see below)
indicates high inclination. For $i \geq 60^\circ$ the expected period is $\geq 22$\,h. 

The H$\alpha$ emission of 2M\,1207 is thought to be dominated by accretion: the broad wings with velocities
of up to $\pm 200$\,km/s can not be explained by magnetic activity, the rotational velocity is comparatively
small and can not be responsible for the broad wings, and 
signatures for winds are very weak \citep{Mohanty05.1,Whelan07.1}. 

The most widely used indicator for identifying accreting 
stars is the equivalent width $W_{\rm EQ}$ of the H$\alpha$ line, where weak emission is
due to chromospheric magnetic activity and strong emission, above 
a certain threshold of $W_{\rm EQ}$, is attributed to accretion. 
However, its dependence on the
continuum level precludes $W_{\rm EQ}$ as a universal measure. To take account of
the declining continuum for later spectral types, the threshold for $W_{\rm EQ}$ that
separates accretors from non-accretors among TTS is usually adapted in discrete
steps from $\sim 5$\,\AA~ for spectral types earlier than M0 to something like 
$20$\,\AA~ for late M-type stars \citep{Martin98.4}. 
Furthermore, 
equivalent width measurements of faint continuum sources such as brown dwarfs are intrinsically associated with 
considerable uncertainties. 
This is due to the low S/N and the presence of overlapping absorption features
that form a `pseudo-continuum' which depends on the spectral resolution. 
To avoid these problems, 
following a suggestion by \cite{White03.1}, the full width
at $10$\,\% of the peak height ($W_{\rm 10\,\%}$) of H$\alpha$ 
has been established as more reliable accretion indicator: 
Objects with $W_{\rm 10\,\%} > 200$\,\AA~ can be considered accretors \citep{Jayawardhana03.2}.

\subsubsection{Profiles and line widths}\label{subsubsect:Ha_profile_width}

\begin{figure}
\begin{center}
\parbox{15cm}{
\parbox{5cm}{\includegraphics[width=5cm, angle=0]{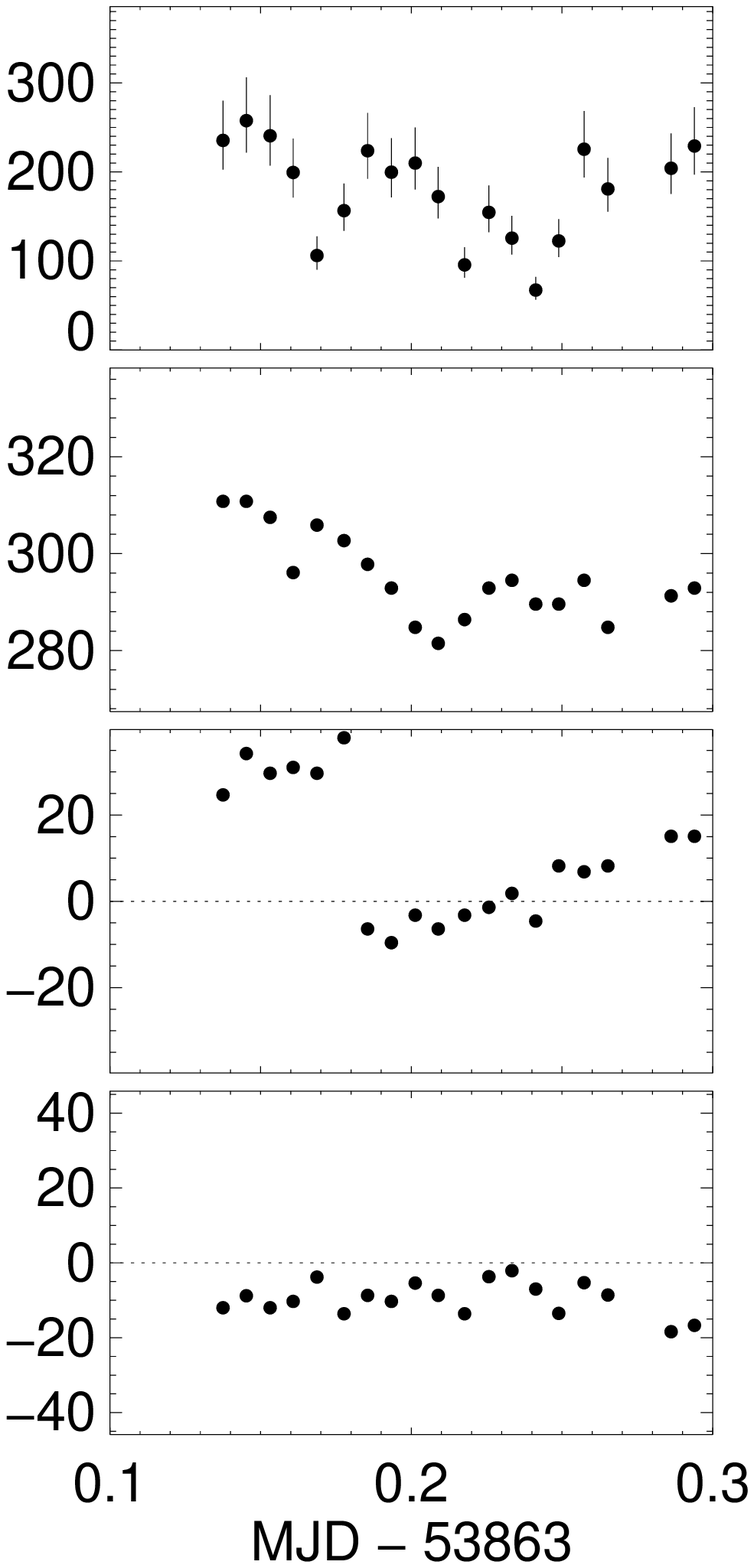}}
\parbox{10cm}{\includegraphics[width=10cm, angle=0]{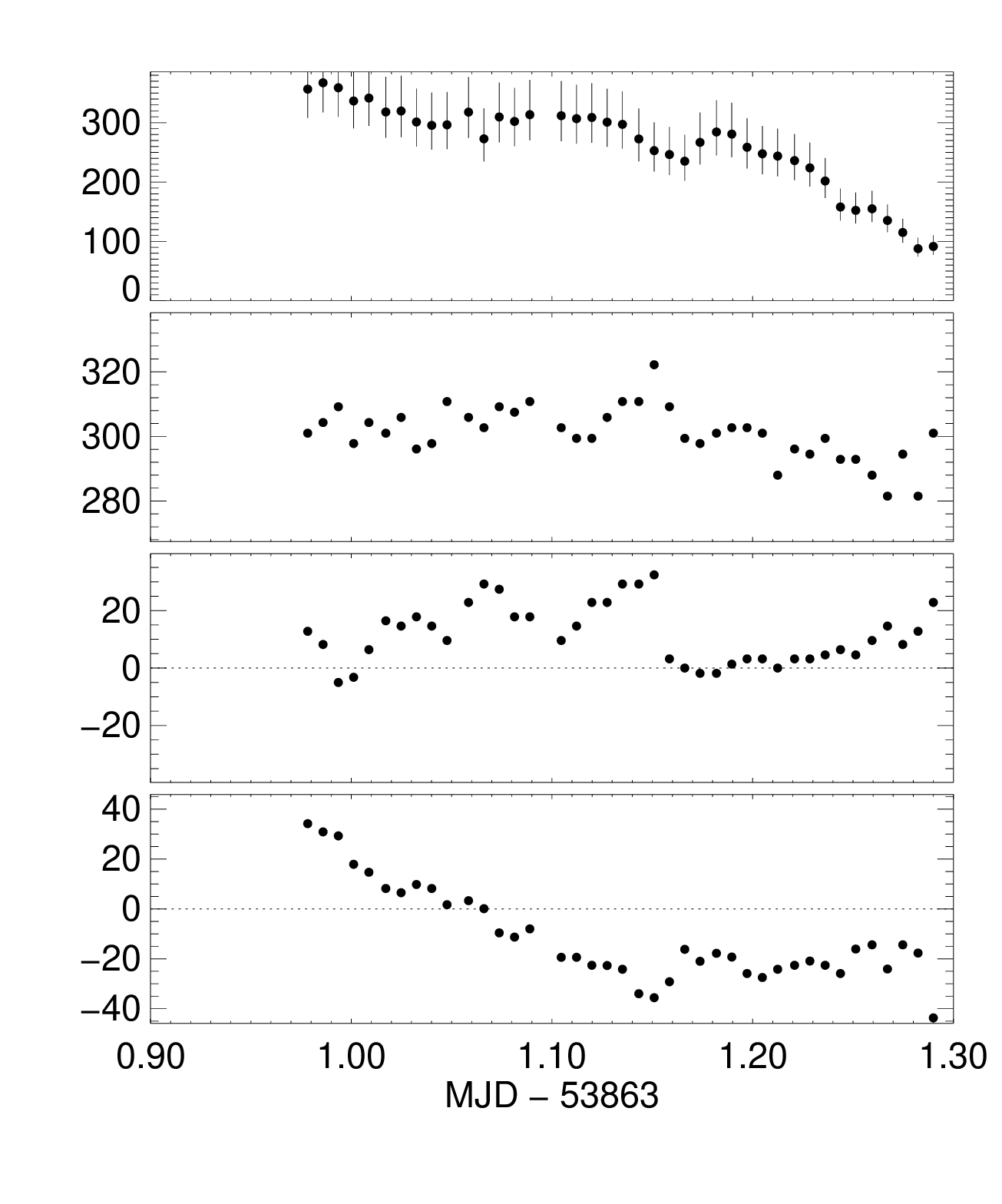}}
}
\caption{Temporal evolution of parameters characterizing the H$\alpha$ emission of 2M\,1207 during May 8-9, 2006.
\label{fig:ha_timeseries}}
\end{center}
\end{figure}

We use $W_{\rm 10\,\%}$ as the primary accretion indicator, but examine also other parameters
that characterize the line intensity and shape. Fig.~\ref{fig:ha_timeseries} shows 
the time-series of these parameters obtained during our run with UVES on May 8-9, 2006. 
The $10$\,\% width is clearly above the accretion threshold 
throughout our observation ($W_{\rm 10\,\%} \approx 280...320$\,km/s); 
on average it is higher during the second night, 
in the course of which it slightly declines. 
The equivalent width has ups and downs during the first night, and 
has its maximum at the beginning of the second night, after which it 
declines dramatically and in a continuous way until the end of the observation. 
The errors in $W_{\rm EQ}$ are dominated by the (faint) continuum level.  
We estimated the uncertainty of the continuum flux by varying the sky extraction area, 
and found that it is of the order of $15$\,\%. Then we computed error bars for $W_{\rm EQ}$
assuming for the continuum flux $0.85 \cdot I_{\rm cont}$ and $1.15 \cdot I_{\rm cont}$, respectively.
Fig.~\ref{fig:ha_timeseries} shows that when taking into account these uncertainties, 
the observed variations of $W_{\rm EQ}$ remain significant. 

\begin{figure}
\begin{center}
\parbox{17cm}{
\parbox{5.5cm}{\includegraphics[width=5.5cm, angle=0]{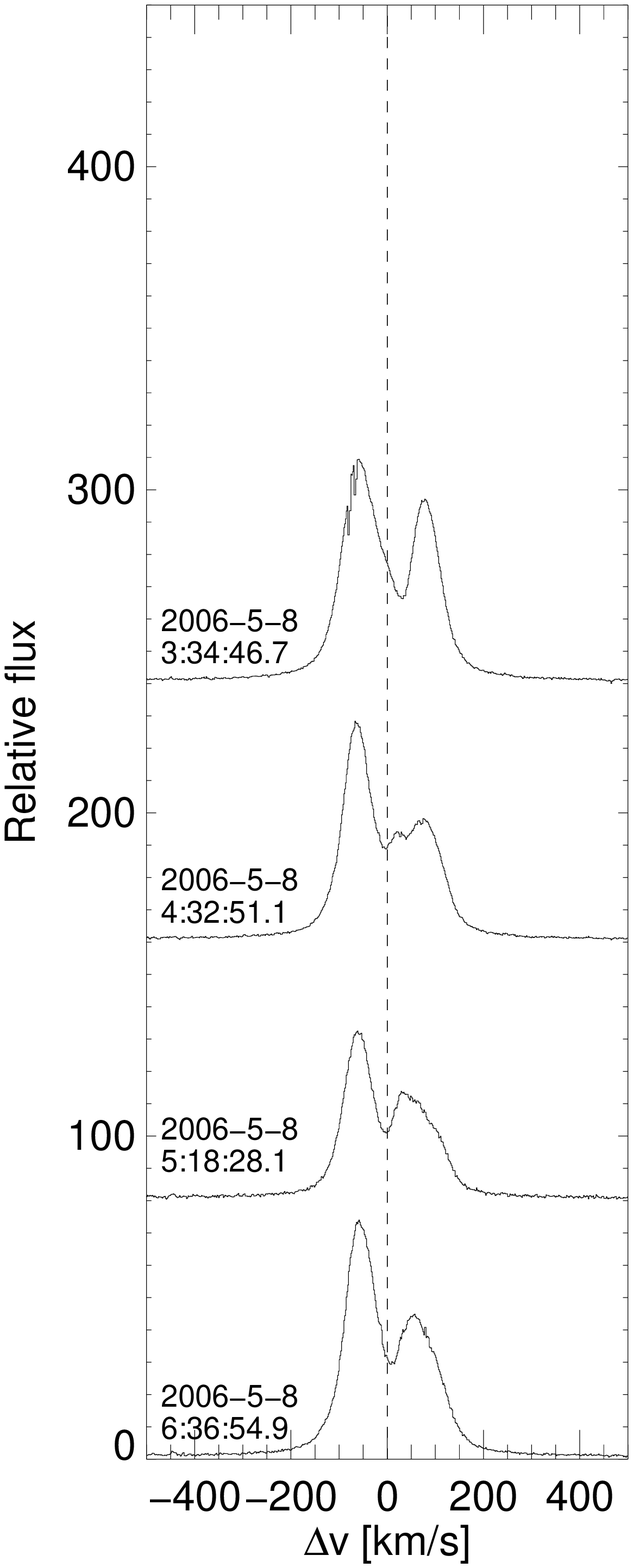}}
\parbox{5.5cm}{\includegraphics[width=5.5cm, angle=0]{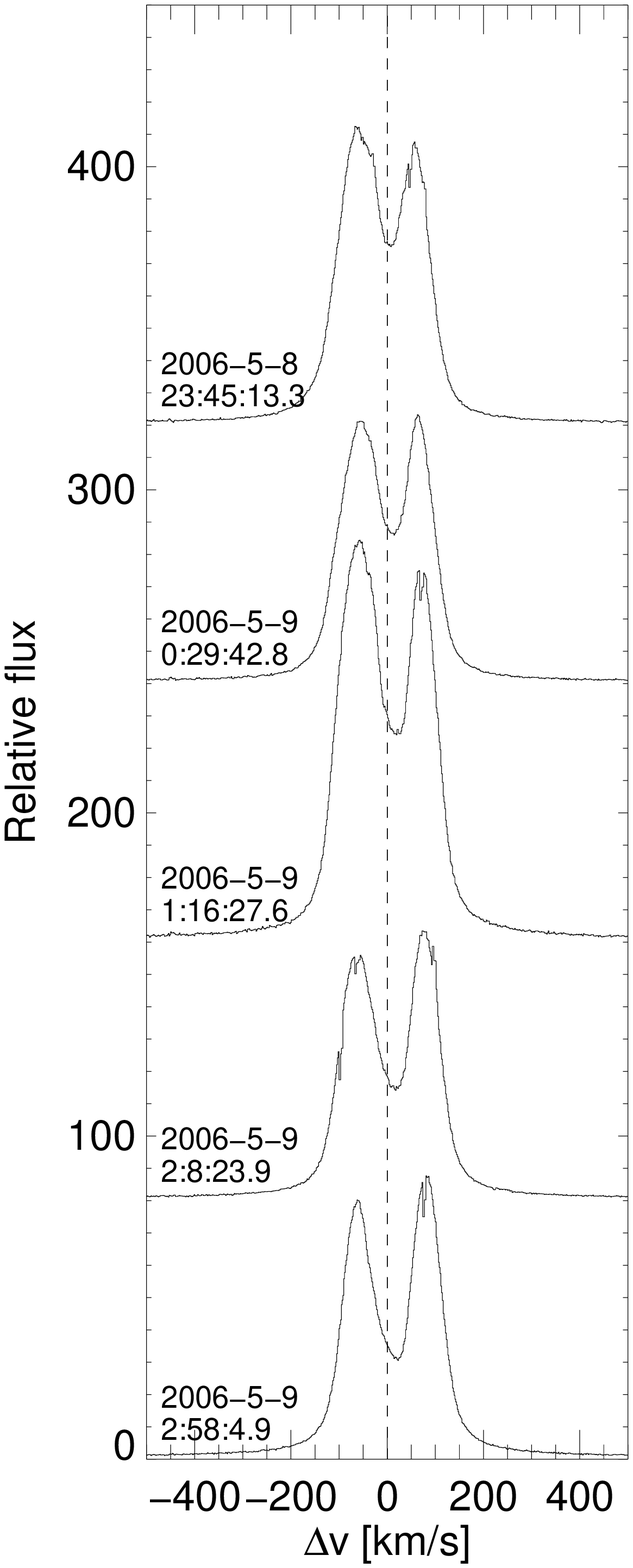}}
\parbox{5.5cm}{\includegraphics[width=5.5cm, angle=0]{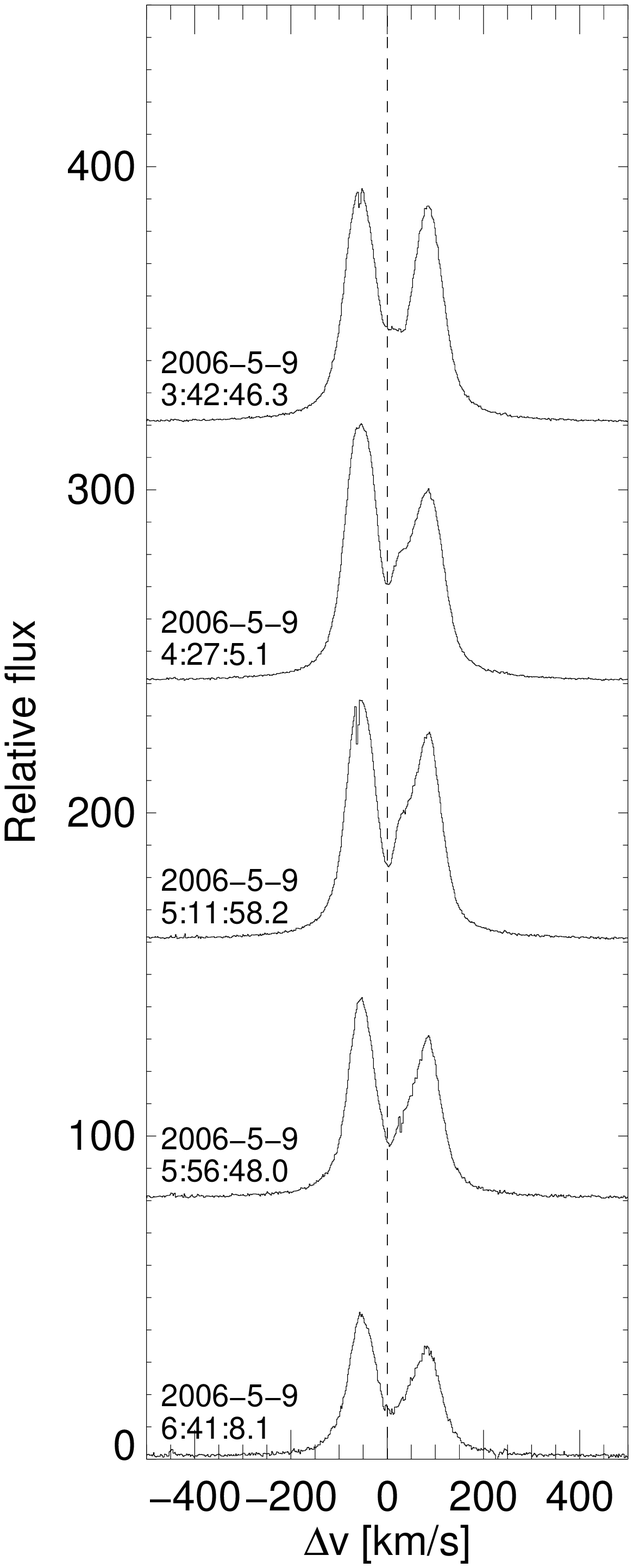}}
}
\parbox{17cm}{
\parbox{5.5cm}{
\includegraphics[width=5.5cm, angle=0]{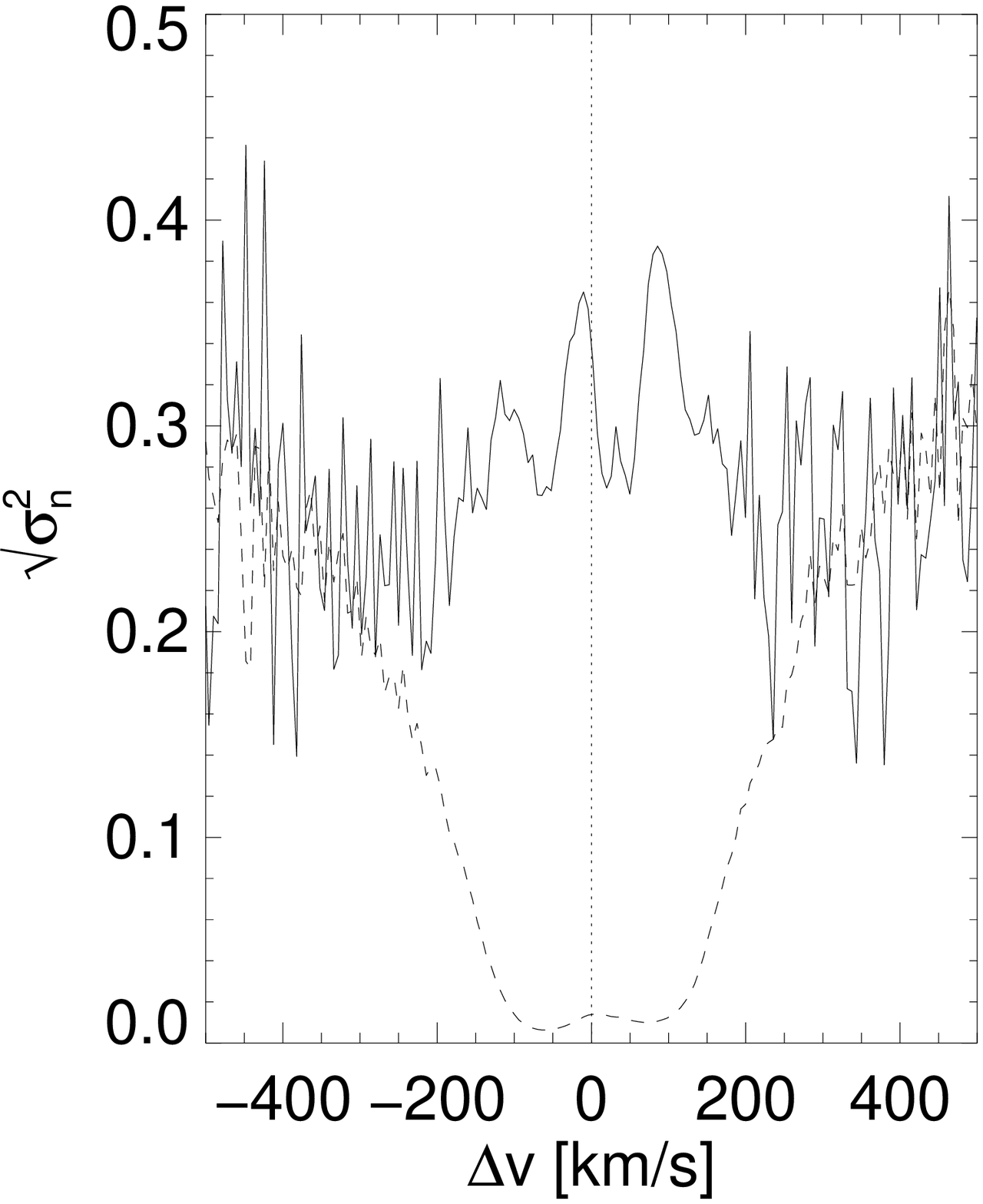}
}
\parbox{5.5cm}{
\includegraphics[width=5.5cm, angle=0]{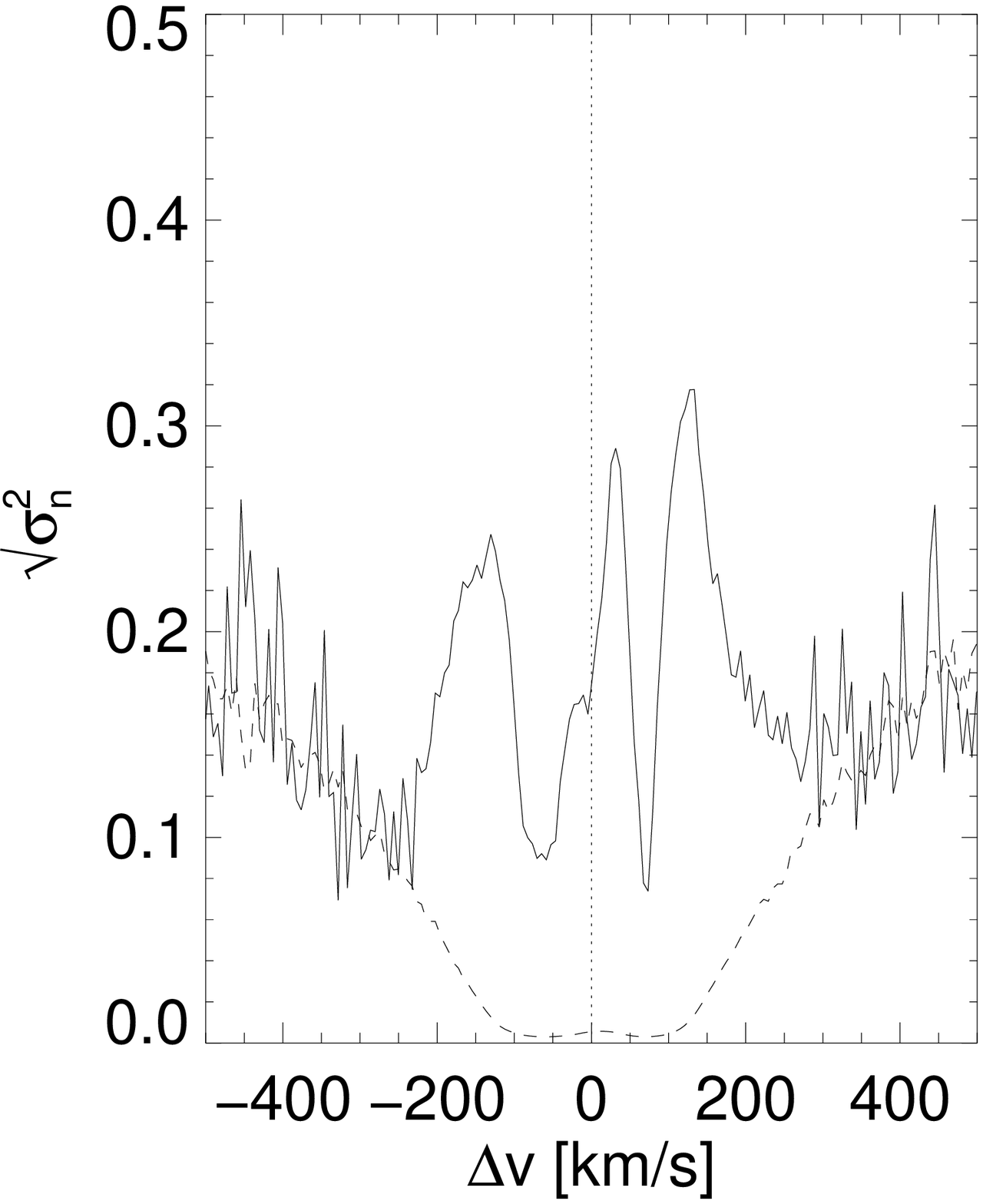}
}
\parbox{5.5cm}{
\includegraphics[width=5.5cm, angle=0]{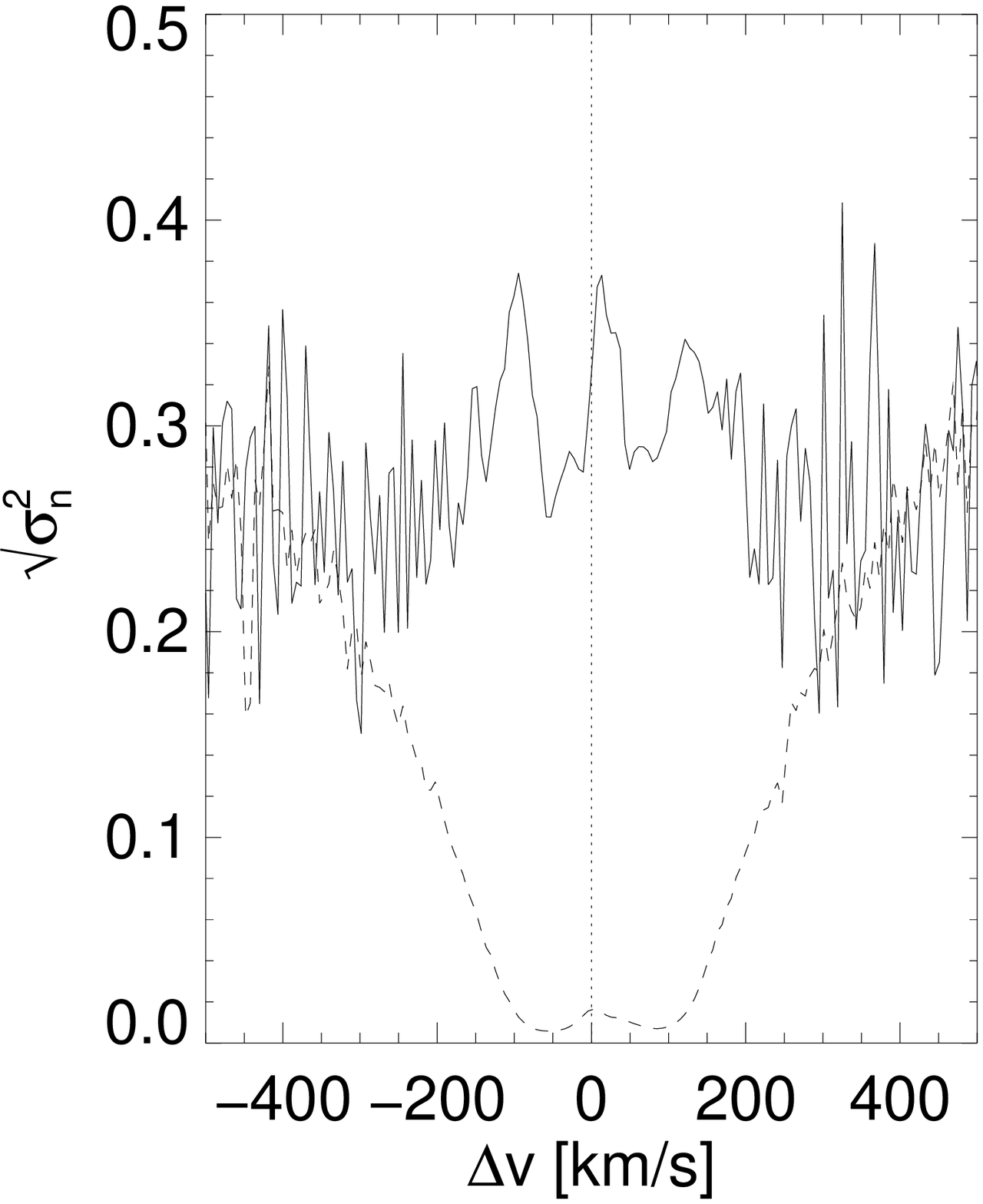}
}
}
\caption{{\em top} -- Normalized average H$\alpha$ line profiles of 2M\,1207 obtained with UVES in May 2006; {\em bottom} - Normalized variance line profiles and zero-variability level.\label{fig:Ha_profiles}}
\end{center}
\end{figure}

\begin{figure}
\begin{center}
\includegraphics[width=5.5cm, angle=0]{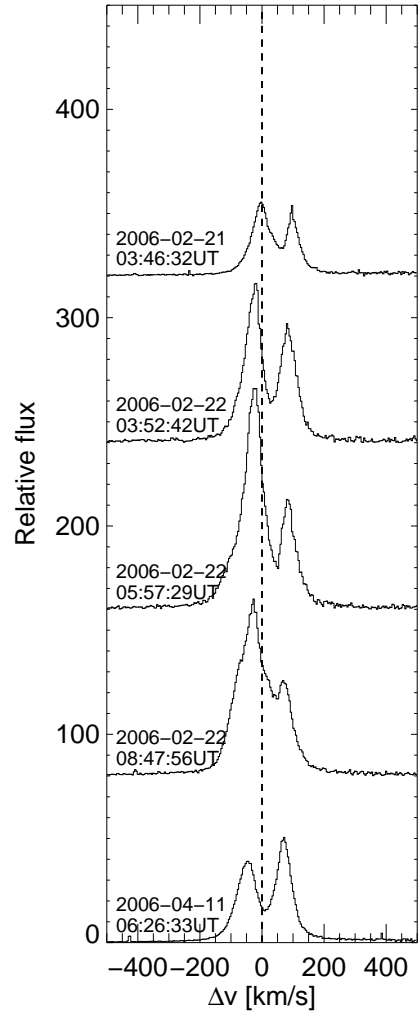}
\caption{H$\alpha$ line profiles of 2M\,1207 obtained with MIKE in Feb and Apr 2006. The data were normalized to the average continuum level near the line. \label{fig:Ha_profiles_mike}}
\end{center}
\end{figure}

The H$\alpha$ emission is double-peaked throughout all our observations from February to May 2006.   
Fig.~\ref{fig:Ha_profiles} displays the time-series of all $14$ normalized average line profiles
$\bar{I_{\rm n}}(\lambda)$ from the UVES run (May 2006), and Fig.~\ref{fig:Ha_profiles_mike} shows the
$5$ normalized profiles obtained with MIKE (February and April 2006).  
For clarity, in Figs.~\ref{fig:Ha_profiles} and~\ref{fig:Ha_profiles_mike} the individual spectra have been 
shifted with respect to each other in the vertical direction. 

\cite{Scholz05.4} have shown that the double-peaked H$\alpha$ profile of 2M\,1207 
can be interpreted as a broad emission line onto which a red-shifted absorption component is superposed. 
For the minimum of the absorption feature we measure a position that varies 
in the range $\Delta v_{\rm min} \approx -10...+70$\,km/s with respect to the expected wavelength
($\Delta v_{\rm min} \approx +30...+70$\,km/s in February 2006, 
$\Delta v_{\rm min} \approx +10$\,km/s in April 2006, and 
$\Delta v_{\rm min} \approx -10...+40$\,km/s in May 2006). 
From Fig.~\ref{fig:ha_timeseries} it appears that twice during the extensive monitoring in May 2006 
$v_{\rm min}$ undergoes a discrete jump from high redshift towards the expected line center. 
These times correspond to line profiles in which the absorption trough is wide and asymmetric,
such that the position of the minimum can not be determined precisely 
(Fig.~\ref{fig:Ha_profiles}).  
As a measure for the asymmetry of the emission component, Fig.~\ref{fig:ha_timeseries} shows the difference between
the $10$\,\% width on the blue (left) and on the red (right) side of $\Delta v=0$.
In the course of the UVES observations, $W_{\rm 10\,\%}(l-r)$ varied from $-40...+40$\,km/s.
In particular, during the second night the line moved systematically from the blue
to the red. This is in contrast to the absorption feature shown above to be redwards of $\Delta v = 0$ 
throughout the observation  
(cf. also Fig.~\ref{fig:Ha_profiles}). 

\begin{figure}
\begin{center}
\includegraphics[width=7cm, angle=0]{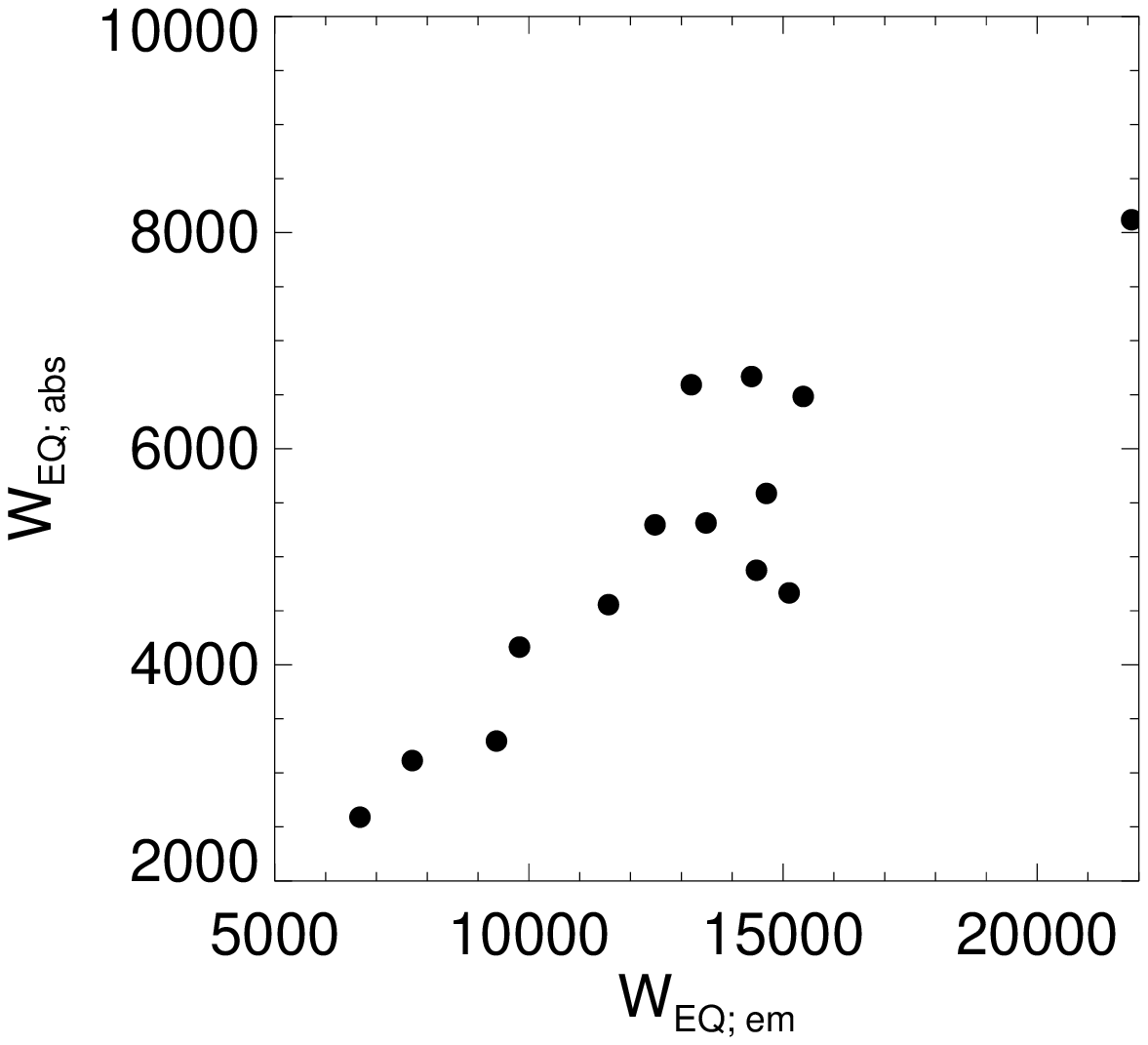}
\includegraphics[width=7cm, angle=0]{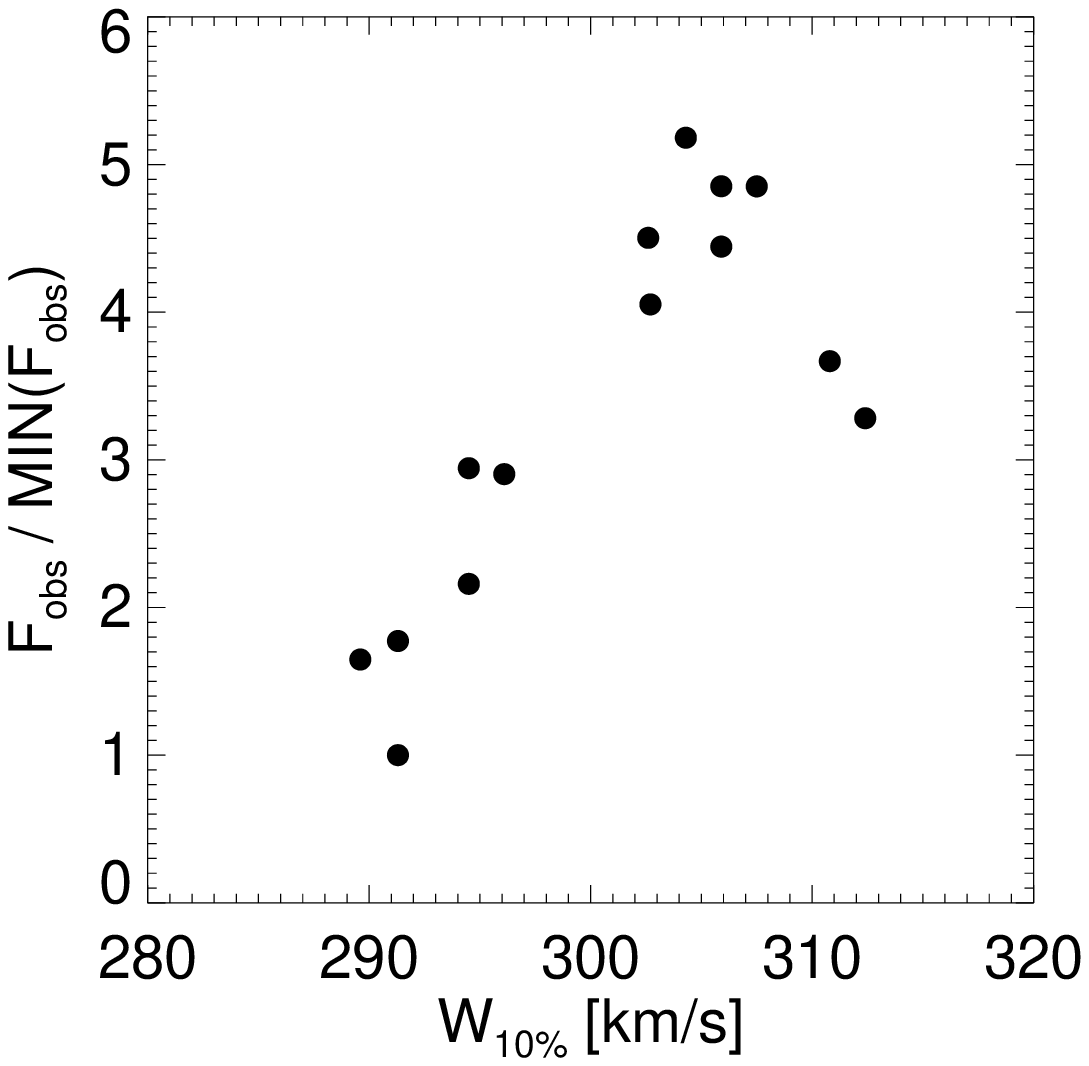}
\caption{Equivalent widths of Gaussian fits to the emission ($W_{\rm EQ; em}$) and absorption ($W_{\rm EQ; abs}$) 
component of the H$\alpha$ line, and normalized total observed flux 
($F_{\rm obs} = F_{\rm em} - F_{\rm abs}$) versus H$\alpha$ $10$\,\% width.\label{fig:Ha_flux}}
\end{center}
\end{figure}

We examined the shape of the H$\alpha$ line and its change in time by 
modeling each of the $14$ average profiles 
with two quasi-Gaussians, 
one with positive normalization representing the emission component and 
one with negative normalization for the absorption component. 
Each of these has three free parameters (position, width, and normalization). 
The equivalent widths of the emission $W_{\rm EQ; em}$ and the absorption $W_{\rm EQ; abs}$ 
component derived from the fits turn out to be strongly correlated (Fig.~\ref{fig:Ha_flux}). 
Since $W_{\rm EQ; em}$ is by definition `normalized' to the continuum this relation is
not trivial.
In Fig.~\ref{fig:Ha_flux} we also plot the {\em observed} flux, i.e. the difference between 
the flux in the emission profile and the absorption profile, versus the $10$\,\% width. 

The H$\alpha$ $10$\,\% width of 2M\,1207 has been measured at eight epochs so far 
(see Table~\ref{tab:w10}). 
In most of the high-resolution spectra collected so far 2M\,1207 
showed $W_{\rm 10\,\%}$ clearly above the accretion threshold, 
although in some occasions there were only marginal signs of accretion 
($W_{\rm 10\,\%} \sim 200$\,km/s).

\subsubsection{Quantifying variability}\label{subsect:var_quant}

To quantify the H$\alpha$ variability as a function of wavelength we examined the 
normalized variance profile
\begin{equation}
\sigma^2_{\rm n}(\lambda) = \sigma^2(\lambda) / \bar{I}_{\rm n}(\lambda),  
\label{eq:varprof}
\end{equation}
where the variance profile 
$\sigma^2(\lambda) = \frac{1}{N-1} \sum_{i=1}^N{[I_{\rm n,i}(\lambda)  - \bar{I}_{\rm n}(\lambda)]^2}$
measures flux variations across the line profile \citep{Johns95.1}.  
In the bottom panels of Fig.\ref{fig:Ha_profiles} we plot $\sqrt{\sigma^2_{\rm n}}$ 
for three distinct parts of the observation in May 2006. From left to right, are shown 
the variances obtained from 
all $19$ spectra from the first night, the first $20$ and the last $20$ spectra of the
second night, respectively. 
This way, the individual $\sqrt{\sigma^2_{\rm n}}$ give information about the variations as a function
of wavelength on short time-scales (within the $\sim 4$\,h covered by $19$ respectively $20$ 
consecutive frames), 
and a comparison of the different $\sqrt{\sigma^2_{\rm n}}$'s probes changes up to $\sim 1$\,d. 

The normalized variance in the continuum expected from Poisson noise is
$\sigma^2_{\rm n, exp}(\lambda) = \sigma^2_{\rm n}(\lambda) (\bar{\sqrt{I}}/\bar{I})^2$. 
The zero-variability level $\sqrt{\sigma^2_{\rm n, exp}}$ 
is shown as dashed line in the bottom panels of Fig.~\ref{fig:Ha_profiles}.
Its comparison with the $\sqrt{\sigma^2_{\rm n}}$'s shows clear signs for variability in the
line profile. 
Throughout the time series, the variance profile (i.e. variability as a function of wavelength)
features three distinct peaks, roughly at positions of $-100...-150$, $-20...+20$, and $+100...+150$\,km/s.
We note that the same structure was seen by \cite{Scholz06.1} in their H$\alpha$ 
time series of 2M\,1207.
The position of these peaks (indicating the velocities of maximum changes in the H$\alpha$ flux)
is roughly coinciding with the position of the two maxima and the minimum in the H$\alpha$
profile. 

\begin{figure}
\begin{center}
\includegraphics[width=15cm, angle=0]{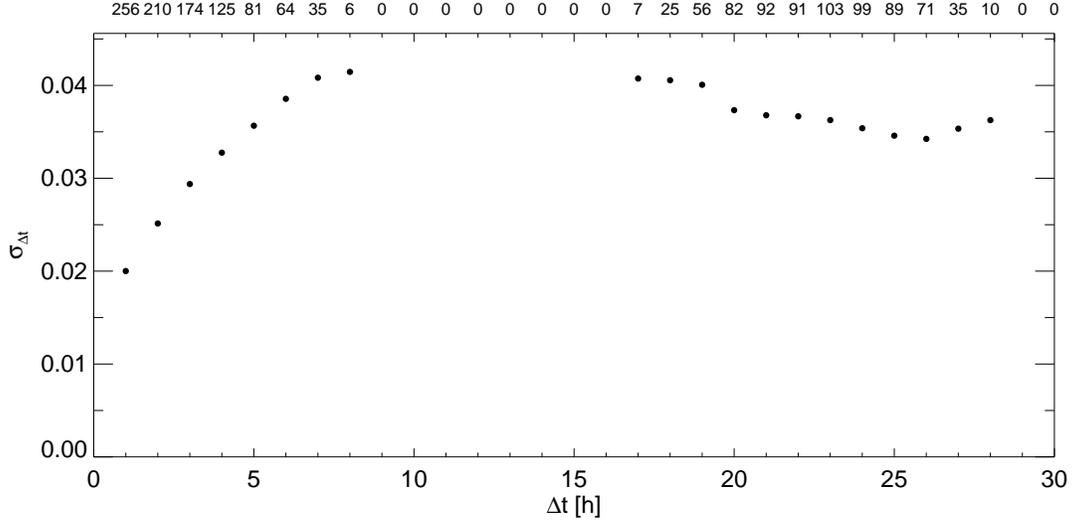}
\caption{Mean of the standard deviation of fractional profile changes $D(\lambda)$ in the H$\alpha$ line as a function of time lag. Numbers on top of the panel indicate the number of data points in each of the time lag bins.\label{fig:ds}}
\end{center}
\end{figure}

The variance profile yields only the time average of the changes as a function of wavelength. 
To study the profile variations in more detail we examine 
the change of the flux between consecutive spectra across the line. 
The relative flux change is given by 
\begin{equation}
D_{\rm ji}(\lambda) = \frac{1}{2} \times \frac{I_{\rm n,j}(\lambda) - I_{\rm n,i}(\lambda)}{I_{\rm n,j}(\lambda) + I_{\rm n,i}(\lambda)},
\label{eq:ampl_change_1}
\end{equation}
where $I_{\rm n}(\lambda)$ is the normalized profile, $i$ is a running number denoting the
exposures and $j = i + 1$. 
These relative flux changes $D(\lambda)$ are used to investigate variations 
as a function of the time lag between two spectra \citep{Johns95.1}. 
To this end, the $D(\lambda)$'s are computed not only for two subsequent spectra but 
for all pairs of the $59$ spectra from May 8-9, 2006, 
i. e. for each time lag $\delta t = t2 - t1 > 0$ sampled by our UVES data. 
Then, we group the $D(\lambda)$'s in time lag bins of $1$\,h (${\rm \Delta t}$) 
and calculate the standard deviation $\sigma_{\rm \Delta t}(\lambda)$ 
of the $D(\lambda)$'s in each of these bins. Finally, we calculate for each time bin 
the average of these standard deviations over the whole line profile ($\bar{\sigma}_{\rm \Delta t}$). 
The result is shown in Fig.~\ref{fig:ds}, where 
the numbers on top of the panel indicate the number of data points in each of the time lag bins. 
The gap is produced by the absence of observations with time lags between
$\approx 9...16$\,h, i.e. the data to the left of this gap corresponds to spectra from 
the same night, and the data to the right corresponds to spectra from two different nights. 
The level of profile variability doubles  
from $\sim 2$\,\% for the shortest time lag ($1$\,h) to   
a maximum of $\approx 4$\,\%, that is reached for a timelag of $\sim 8$\,h.

To test whether the variations are correlated across the line profile we computed the correlation
coefficient for each pair of spectral bins using the profiles $\bar{I}_{\rm i}(\lambda)$,  
with $i=1,...,N$. 
This results in a correlation matrix $r_{\rm i,j}$ 
of dimension $N_{\rm b} \times N_{\rm b}$, where $N_{\rm b}$
is the number of spectral bins. The confidence level of each coefficient in the matrix is computed
using the error function 
\begin{equation}
E_{\rm i,j} = \frac{\mid r_{\rm i,j} \mid \sqrt{N_{\rm b}}}{\sqrt{2}},
\label{eqw:erf}
\end{equation}
such that $1-E_{\rm i,j}$ is the confidence level for the correlation to be true. 

\begin{figure}
\begin{center}
\parbox{17cm}{
\parbox{5.5cm}{
\includegraphics[width=5.cm, angle=0]{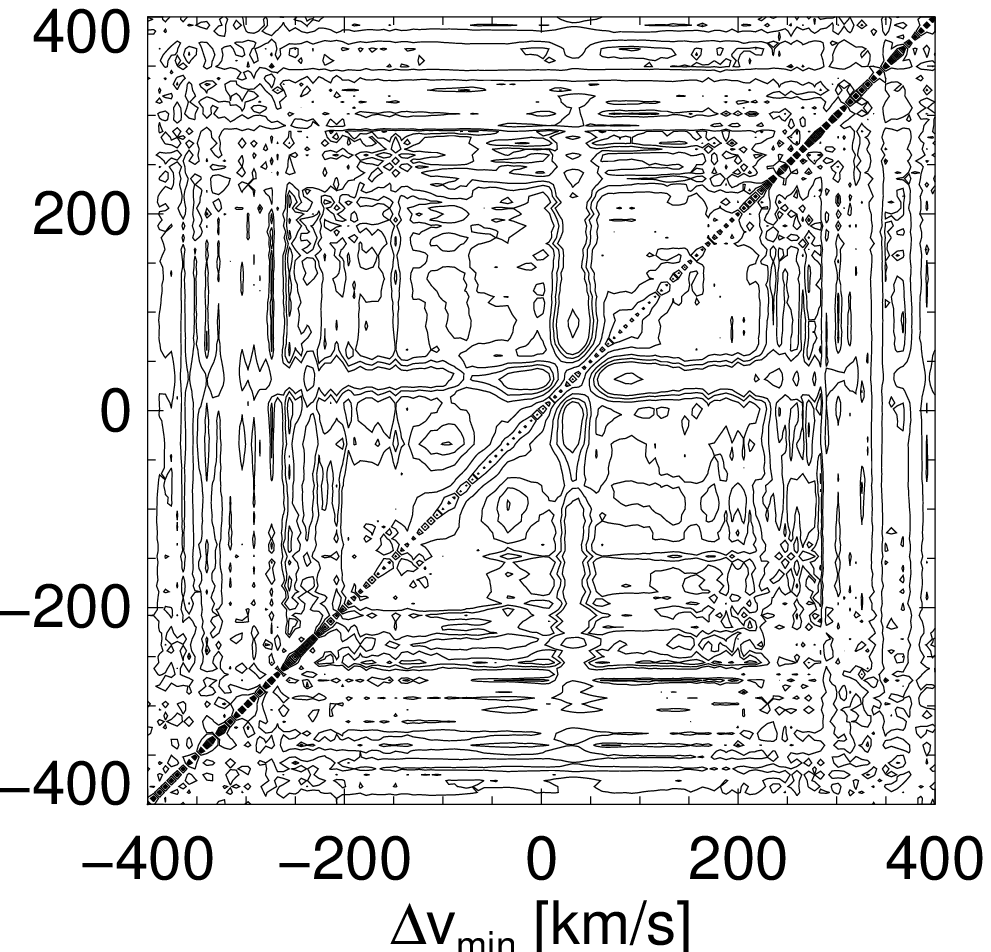}
}
\parbox{5.5cm}{
\includegraphics[width=5.cm, angle=0]{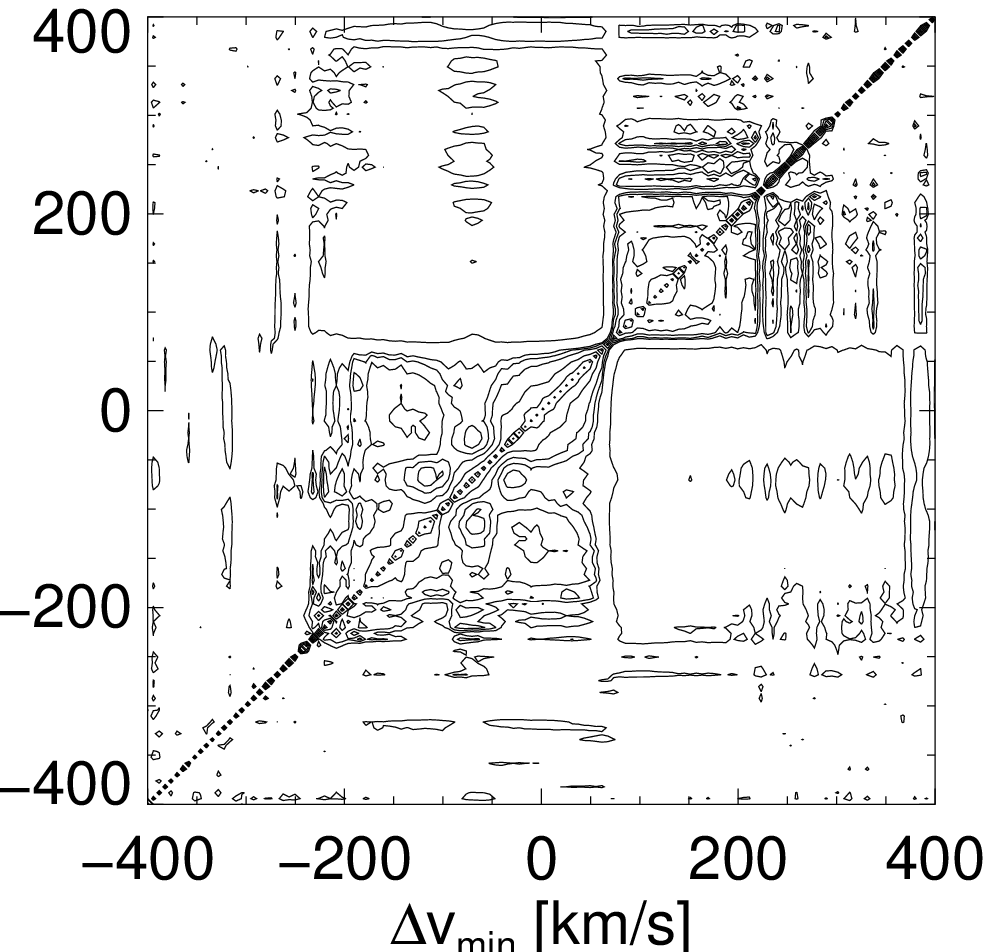}
}
\parbox{5.5cm}{
\includegraphics[width=5.cm, angle=0]{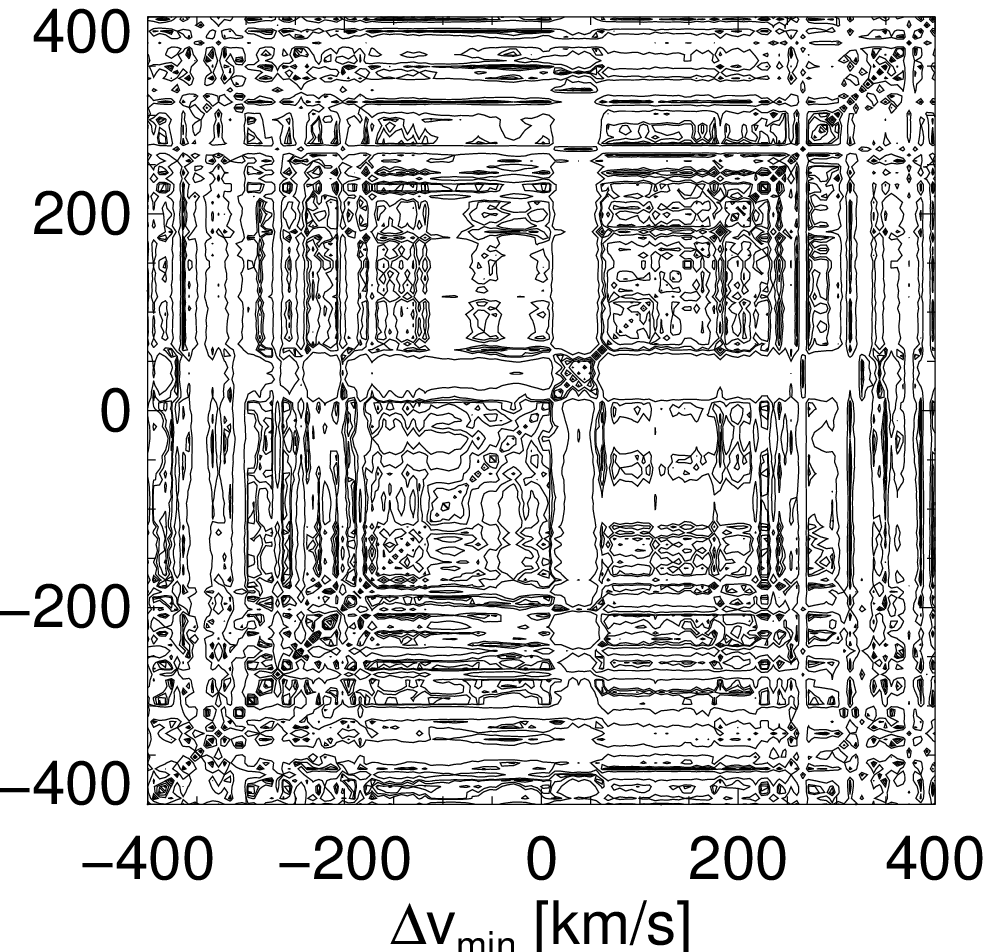}
}
}
\caption{Correlation matrix of H$\alpha$ for different subset of exposures (denoted $i$) obtained on May 8-9, 2006.
{\em left} -- $i=1,...,19$; {\em middle} -- $i=20,...38$; {\em right} -- $i=7,...,10$.}
\label{fig:corr_matrix_Ha}
\end{center}
\end{figure}

%
The correlation matrix has been computed for different subgroups of H$\alpha$ profiles from 
the UVES run. Fig.~\ref{fig:corr_matrix_Ha} displays the most interesting correlation matrices
$r_{\rm i,j}$ as contour plots. Only values with
confidence level $>99.9$\,\% are considered in this representation. The first two
diagrams from the left represent the matrices of each $19$ consecutive spectra, obtained in
the first and the second observing night, respectively. The cross-shaped pattern seen in the 
leftmost diagram
is peculiar. The center of the cross is offset by $\approx +25$\,km/s to the red side, and it
represents the spectral region near the minimum of the line profile (see Fig.~\ref{fig:ha_timeseries}). 
Closer examination shows that this pattern stems from a short time interval. 
This is demonstrated in the rightmost diagram of Fig.~\ref{fig:corr_matrix_Ha}, 
that displays the correlation matrix for exposures $i = 7,...,10$, 
roughly corresponding to the second average profile in the leftmost panel of Fig.~\ref{fig:Ha_profiles}.
Clearly, during this time interval, the profile changes near the 
absorption minimum are uncorrelated with all velocities except for the velocities in close vicinity. 
The double-square shaped pattern observed during the second night 
(middle panel of Fig.~\ref{fig:corr_matrix_Ha}) is more typical \citep[cf.][]{Johns95.1}. 
The absence of contours in the upper left and in the lower right
of this graph means that red and blue wings vary in an uncorrelated fashion. The range near
$v \approx +75$\,km/s, corresponding to the position of the right emission peak, 
is particularly well correlated with the whole line.

\subsection{Other signatures of accretion}\label{subsect:acc_lines}

Our high S/N UVES spectra allow us to present the first time series for other optical emission
lines of 2M\,1207. 
In particular, we detect the Balmer lines up to H$\epsilon$. This latter one 
is partly blended with Ca\,II\,H\,$\lambda3968$.

\begin{figure}
\begin{center}
\parbox{17cm}{
\parbox{8cm}{
\includegraphics[width=7cm, angle=0]{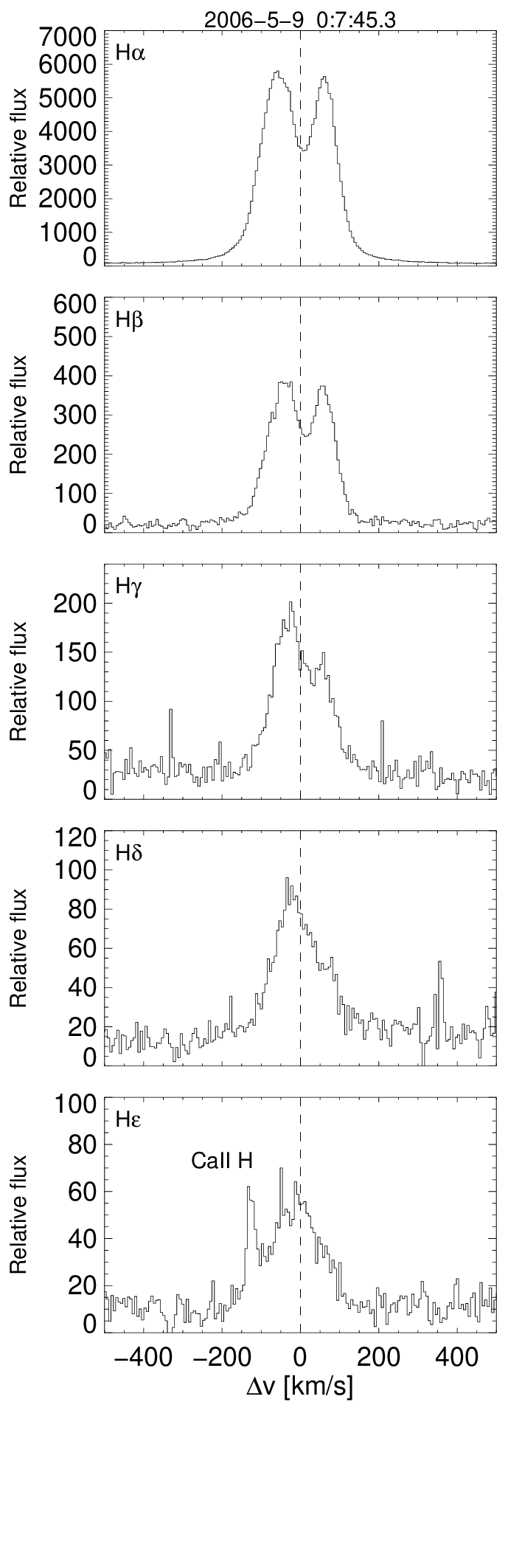}
}
\parbox{8cm}{
\includegraphics[width=7cm, angle=0]{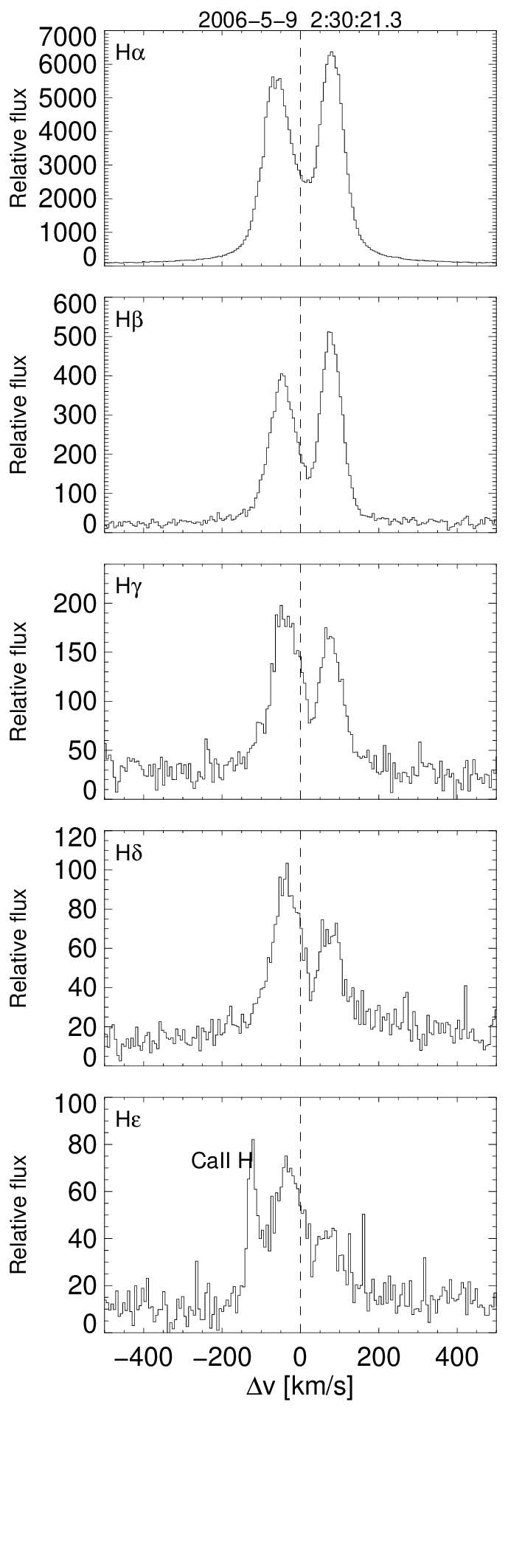}
}
}
\caption{Contemporaneous Balmer emission line profiles (H$\alpha$ to H$\epsilon$) for two epochs during the UVES run (8-frame averages). \label{fig:balmer}}
\end{center}
\end{figure}

The shape of the H$\beta$ profile
follows closely that of H$\alpha$ discussed in detail above. All higher Balmer lines are predominantly 
characterized by a double-peaked profile, but at times turn into a single-peaked shape. 
This is demonstrated in Fig.~\ref{fig:balmer}, that shows the contemporaneous Balmer series
for two epochs. 
Each panel in Fig.~\ref{fig:balmer} is obtained from an $8$-frame average. 
When present, the absorption minimum is found at roughly the same position for all Balmer lines, 
but the ratio of the emission fluxes left and right of the minimum decreases
systematically for the higher Balmer lines, i.e. the lower Balmer lines are red-dominated and 
the higher Balmer lines are blue-dominated. 

In contrast to the Balmer lines, the Ca\,II H+K emission is narrow with FWHM of only $\sim 20$\,km/s 
(see Table~\ref{tab:eqw}). 
Furthermore, we detect several lines of He\,I ($\lambda4471$, $\lambda5876$, $\lambda6678$, $\lambda7066$) 
and the Ca\,II infrared triplet (hereafter Ca\,IRT) in emission. All these lines
are predominantly detected in accretors and, therefore, thought to be related to the accretion process. 
The Ca\,II\,$\lambda8662$ flux was shown to be directly related to the mass 
accretion rate \citep{Mohanty05.1}. 
The three lines of the Ca IRT appear with roughly similar strength,
inconsistent with formation in an outflow \citep{Reipurth86.1, Fernandez01.1}. 
No Na\,D emission is seen, consistent with the prediction of magnetospheric accretion models 
for low accretion rates \citep{Muzerolle01.1}. 

We measured $W_{\rm EQ}$ and the FWHM of the stronger ones of the emission lines discussed above. 
The continuum was estimated from two $5$\,\AA~ wide line-free 
regions left and right of the respective line. 
The observed range of $W_{\rm EQ}$ is reported for each line in Table~\ref{tab:eqw} together with
its approximate uncertainty. Given the often loose use of equivalent widths in the literature, 
it seems appropriate to stress that $W_{\rm EQ}$ is associated with considerable errors in 
objects with faint continuum. This is clear from the numbers given in Table~\ref{tab:eqw}. 
The errors cited in that table do not include systematic uncertainties, again affecting 
mainly the (faint) continuum, that we discovered in the process of defining the sky extraction 
area. Since we are interested in relative changes within the UVES dataset, systematic 
uncertainties are not a big concern. However, they should be kept in mind when comparing data 
obtained with different instruments, and analysed in different ways by different astronomers. 

\begin{figure}
\begin{center}
\includegraphics[width=8cm, angle=0]{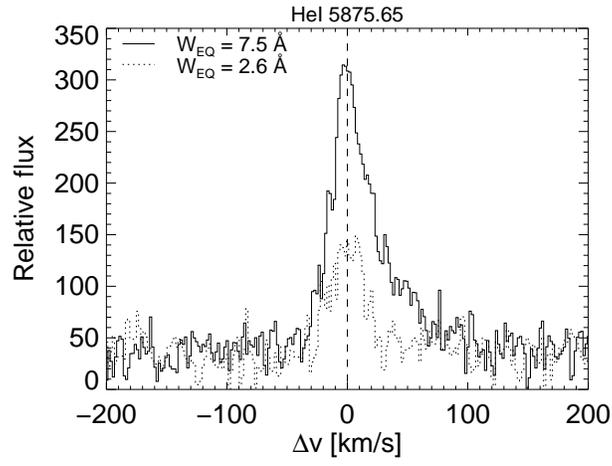}
\caption{Two spectra in the region of He\,I\,$\lambda5876$ showing the change from the typical profile with extended red wing (solid line) to the weak, symmetric line (dotted line).\label{fig:heI}}
\end{center}
\end{figure}

All emission lines are stronger in the second night of the UVES observations, and they are particularly
weak towards the end of the first night (around MJD\,53863.25  = UT 05\,h on May 8, 2006). 
At the same moment the shape of some of the strongest emission lines
underwent a change: 
(i) He\,I\,$\lambda5876$ temporarily becomes flat-topped and symmetrical, while
it is markedly asymmetric with an extended red wing throughout most of the two nights 
(see Fig.~\ref{fig:heI}); 
(ii) in H$\alpha$ the $10$\,\% width is at its minimum, and 
the deficiency in the red vs. the blue peak is particularly strong; 
(iii) in the higher Balmer lines the red portion of the line disappears almost completely, 
and (iv) some of the weaker lines nearly disappear at the same time.

\section{Discussion}\label{sect:discussion}

\subsection{Spectral variability}\label{subsect:specvar}

We have examined the optical emission line variability of 2M\,1207 at high spectral
resolution 
in data obtained at various epochs through February, April, and May 2006.
In particular, we present for the first time a near continuous spectral sequence
from two consecutive nights with a time-resolution of $\approx 12$\,minutes. 
The study presented here complements previous work that was characterized by much poorer
temporal sampling.

\subsubsection{Short timescales (days to hours)}\label{subsubsect:short}

\citet{Scholz05.4} have attributed the drastic changes of the H$\alpha$ line profile seen in 
earlier high-resolution spectra of 2M\,1207 to varying viewing angle of the
accretion column(s) over the course of the brown dwarf's rotation cycle. The absorption component
seemed to appear and reappear on a time-scale of $\sim 1$\,day, 
an interval approximately coincident with the expected rotation period, 
therefore suggesting that the accretion spots moved with respect to the observer. 
However, with a maximum of $3$ spectra per night, 
the available data did not provide adequate sampling across the presumed rotation cycle. 

Within the two nights of near-continuous observation in May 2006 
there is clear variability present in the H$\alpha$ emission and other emission lines, 
albeit much less pronounced and systematic than expected from previous data. 
Here, we summarize its characteristics: 
(i) The line profile is double-peaked throughout the two nights, displaying the central
absorption reversal characteristic for accretion in star-disk systems seen near edge-on; 
(ii) the absorption feature moves between $-10...+40$\,km/s; 
(iii) the absorption minimum widens twice during the observations with a time interval of
$\sim 23$\,h;
(iv) variability across the line is inferred from the variance profile and the correlation matrix; and 
(v) the line flux varies by a factor of $\approx 6$, 
correlated with the (small) variations of the $10$\,\% width.

The variable redshift and width of the absorption minimum implies that 
the viewing angle of the absorbing layers of the accretion column changed throughout the observation. 
\cite{Bouvier03.1} explained the radial velocities of the redshifted absorption components of the
TTS AA\,Tau with changes in the structure of the magnetosphere. 
In this scenario, during times of high observed
absorption redshift an observer at high inclination looks flat onto the magnetic field lines
that carry the absorbing material,
while low absorption redshift implies a large angle between the absorbing material and the line-of-sight.
The different viewing angles can be realized by inflating or deflating the star-disk field. 
In the case of 2M\,1207, the observed changes that occur on time-scales shorter than the rotation period
may suggest that the magnetosphere is somewhat inhomogeneous across stellar longitude. 
However, the changes are not as dramatic as seen before. 
The range of observed redshifts in the absorption minimum of 2M\,1207 is comparable to that of AA\,Tau
but there is no blueshifted absorption. Therefore, for 2M\,1207 outflows do not play a significant
role in the formation of the H$\alpha$ line. 

Although there is no obvious periodicity, the rotation of the star is the most
likely cause of the short-term changes of the H$\alpha$ profile summarized above.
We have shown that the fractional profile changes reach a maximum after a time-lag of $\sim 8$\,h,
and they remain high on a plateau for larger timelags. (The exact position of the maximum is
unclear due to the absence of data with timelags between $9-15$\,h.) This timescale is roughly 
consistent with one half the expected rotation period, and therefore with the time interval in which
maximum changes due to rotational effects are expected. 

If possible contributions from chromospheric activity and winds are neglected, 
the emission line flux is expected to be a measure for the accretion rate, and this is supported
by its correlation with the $10$\,\% width. 
This correlation is not trivial because the fit function's
normalization is independent of its width, unlike in a `canonical' Gaussian. 
We conclude, from the correlation observed between the equivalent widths of the emission and
the absorption components  
(Fig.~\ref{fig:Ha_flux}) 
that the amount of absorbing material is proportional to the total amount of material crossing 
the line-of-sight. 

With the exception of the Ca\,II\,H + K lines, the strong emission lines 
(higher Balmer lines, He\,I and Ca\,IRT) follow the H$\alpha$ emission closely 
in their shape and variability, indicating a common origin, 
i.e. accretion, and chromospheric contributions are not significant. 
In particular, throughout our observations the lines of the Balmer series 
are mostly double-peaked. Only the higher lines H$\delta$ and H$\epsilon$ at times turn into a single-peaked shape.

\subsubsection{Intermediate timescales (weeks to months)}\label{subsubsect:intermediate}

The H$\alpha$ line is characterized by a double-peaked profile in all spectra obtained
from February to May 2006. In the spectra from February and April 
the redshift of the absorption feature is systematically larger with respect to
the observations from May of the same year, and consequently  
the line profiles observed in February and April are also more asymmetric
than the ones from May. However, in the data from 2006 
the variations are smoother and less dramatic than the ones from 2005, 
for which \cite{Scholz05.4} reported alternating single- and double-peaked profiles 
within individual observing nights.  

If the previous assumption of a high-inclination system rotating at a rate of 
$\sim 1$\,d is correct, the 
data from 2006 are compatible with a rather homogeneous distribution of accretion elements,
i.e. magnetosphere, across the (sub)stellar longitudes, with smaller inhomogeneities as discussed
in Sect.~\ref{subsubsect:short}. In contrast, during 2005 the absorption reversal appeared
and disappeared within one rotation cycle, implying strong variations in the accretion structure
across the stellar surface. 
We conclude that the distribution of the accretion elements or the structure of the 
magnetosphere has changed within $\sim 1$\,yr. 
In both years, 2005 and 2006, the mass accretion rate, 
as inferred from the $10$\,\% width of the H$\alpha$ emission 
\citep[see e.g.][ and Sect.~\ref{subsect:macc}]{Natta04.2},
has shown fluctuations by a factor $\sim 2$ 
on time-scales of weeks to months (cf. Table~\ref{tab:w10}).

\subsubsection{Long timescales (years)}\label{subsubsect:long}

Three years have passed since the first high-resolution
spectrum of 2M\,1207 was obtained, and in this period there seems to be a tendency  
towards enhanced mass accretion rate. 
Note, however, that the large H$\alpha$ equivalent width measured by 
\cite{Gizis02.2} in a low-resolution spectrum counteracts this trend, and 
fluctuations on shorter timescales (discussed above) may be superposed on the long-term evolution. 
Further observations are
required to confirm our speculation about the secular evolution of the H$\alpha$ emission.

\subsection{Accretion rate and magnetic field}\label{subsect:macc}

\cite{Natta04.2} have parametrized the accretion rate as a function of $W_{\rm 10\,\%}$ for TTS 
and brown dwarfs. 
From their Eq.~1, we find an accretion rate of $10^{-10.1...-9.8}\,{\rm M_\odot/yr}$ 
for the range of H$\alpha$ $10$\,\% widths measured for 2M\,1207 during the UVES run in May 2006. 
In previous measurements, $\dot{M}_{\rm acc}$ was as low as $\sim 10^{-11.2}\,{\rm M_\odot/yr}$,  
i.e. 2M\,1207 changes its accretion rate by a factor of $2$ on the timescale of
days, and at least by one order of magnitude over the years. 

\begin{figure}
\begin{center}
\includegraphics[width=15cm, angle=0]{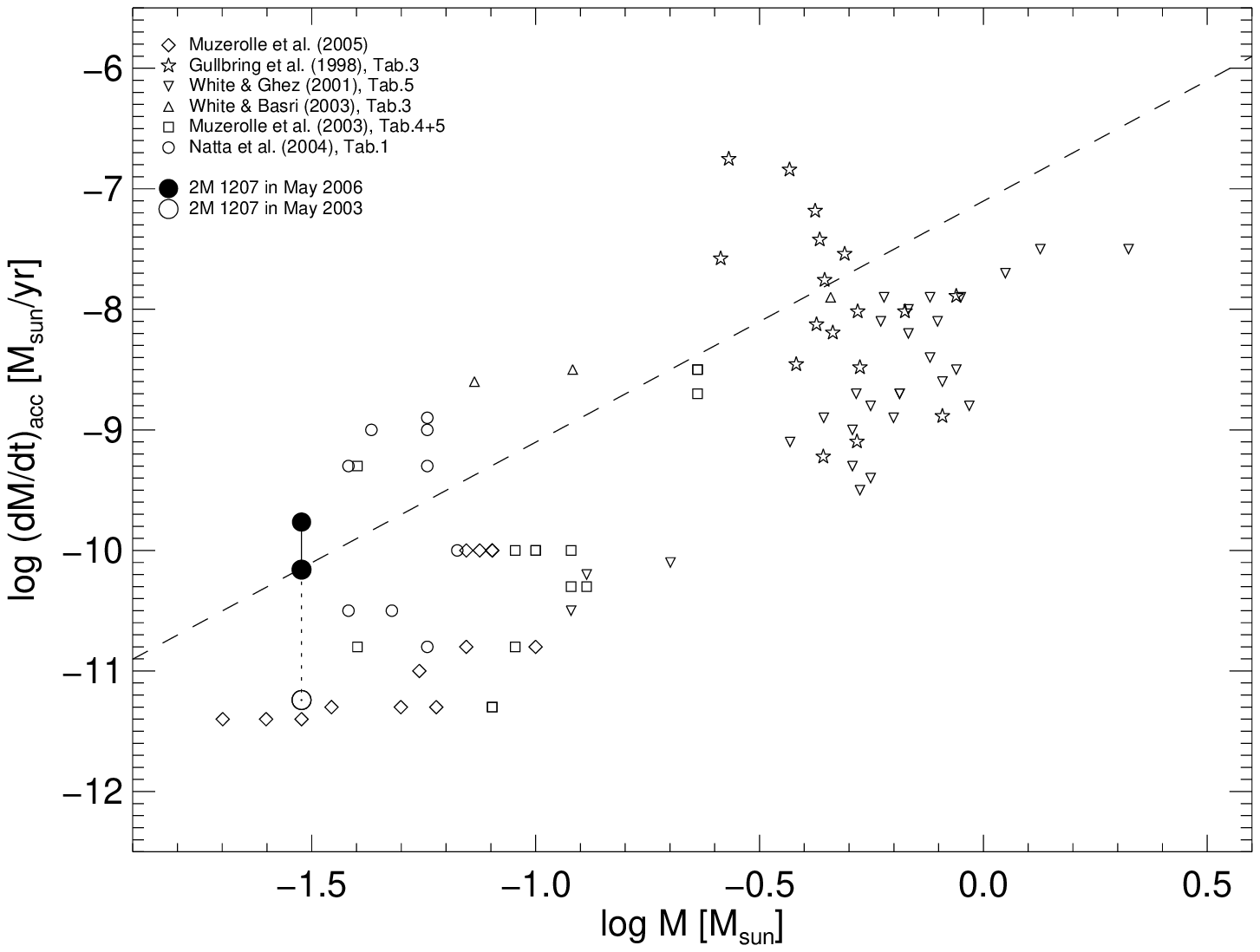}
\caption{Relation between mass and mass accretion rate for sub-stellar objects; data from the literature cited by \protect\cite{Muzerolle05.1}\label{fig:mdot}}
\end{center}
\end{figure}

In Fig.~\ref{fig:mdot} we put 2M\,1207 in the context with other accreting (sub)stellar objects.
A correlation between accretion rate and mass 
($\dot{M}_{\rm acc} \sim M^\alpha$ with $\alpha \approx 2$) has been reported in the literature.
In Fig.~\ref{fig:mdot} we display this relation using data from 
the references given in \cite{Muzerolle05.1}.
The range of $\dot{M}_{\rm acc}$ inferred from all available data of 2M\,1207 is 
indicated by large circles connected by a vertical line. 

Various explanations have been proposed for the correlation between $\dot{M}_{\rm acc}$ and $M$,  
and its large ($>2$\,dex) scatter. 
\cite{Padoan05.1} see in the $\dot{M}_{\rm acc} - M$ relation
a consequence of large-scale Bondi-Hoyle accretion. In their scenario, 
the scatter is produced by the different physical 
conditions in different star forming environments. \cite{Alexander06.1} assumed that the $\dot{M}_{\rm acc} - M$ 
relation reflects the initial parameters of the disk-star system, and they showed that the scatter 
can be reproduced by a combination of varying disk initial parameters and 
the viscous evolution of the disk in the course of which the mass accretion rate decreases. 
According to \cite{Gregory06.1} the $\dot{M}_{\rm acc} - M$ 
relation arises naturally from models of magnetospheric accretion assuming a realistic field
geometry. 
Finally, \cite{Mohanty05.1} suggested declining disk ionization with decreasing 
stellar mass as the origin of the $\dot{M}_{\rm acc} - M$ relation,
implying that most of the scatter is due to intrinsic variability in accretion rates. 
This view was disputed by \cite{Natta06.1} who did not find strong variations of the 
near-IR emission lines indicative of accretion 
in a sample of $14$ objects from the $\rho$\,Oph star forming region.
On the other hand, it has been demonstrated
for a few examples of individual objects, that indeed accretion rates do vary substantially
\citep{Gullbring96.2,Alencar05.1,Scholz05.4,Scholz06.1}. 
We point out that the range of mass accretion rates derived for 2M\,1207 
is comparable to the spread in the $\dot{M}_{\rm acc} - M$ relation. 
Therefore, variability might indeed explain this spread. 
Repeated measurements of $\dot{M}_{\rm acc}$ for known accretors will be helpful to verify our suggestion. 

An alternative method to derive accretion rates has been proposed by \cite{Mohanty05.1}.
They established that both classical TTS and brown dwarfs follow 
a linear correlation between the Ca\,II\,$\lambda8662$ emission flux and
$\dot{M}_{\rm acc}$ determined independently from H$\alpha$ modeling or veiling measurements. 
Following their recipe, we estimate the continuum flux $F_{\rm cont}$ 
of 2M\,1207 at $8662$\,\AA~ from the synthetic 
DUSTY spectra of \cite{Allard00.1}. We adopt $T_{\rm eff} = 2800$\,K, $\log{g} = 4.0$, 
and negligible veiling \citep[see discussion in][ for justification of these choices]{Mohanty05.1}. 
The line flux is then given by $F_{\rm CaII} = F_{\rm cont} \cdot W_{\rm EQ}$.
For the highest equivalent width measured in May 2006 this results in 
$\log{F_{\rm CaII}}\,[{\rm erg/cm^2/s}] \approx 5.1$, 
and the fit relation for the low-mass sample of \cite{Mohanty05.1} yields 
an accretion rate of $\dot{M}_{\rm acc} \approx 10^{-10.3}\,{\rm M_\odot/yr}$. This is in rough
agreement with the value we derived above from the H$\alpha$ $10$\,\% width. We caution, that using 
the Ca\,II\,$\lambda8662$ line for estimating the mass accretion rate assumes a negligible contribution 
from activity. This seems to be true for brown dwarfs on a statistical basis 
\citep[`the $\lambda8662$ component occurs almost exclusively in accreting objects'; ][]{Mohanty05.1},
and for 2M\,1207, in particular, by the lack of measureable effects from chromospheric emission
in other lines.

According to \cite{Koenigl91.1} the mass accretion rate of TTS 
is related to their magnetic field strength,
$B \sim \dot{M}^{1/2} \cdot M_*^{1/4} \cdot R_*^{-3} \cdot R_{\rm t}^{7/4}$. 
We assume $R_{\rm t} \sim 2 R_*$ for the disk truncation radius \citep{Muzerolle00.2}. The mass and
radius of 2M\,1207 are $M_* = 0.03\,M_\odot$ and $R_* = 0.27 R_\odot$ \citep{Chabrier00.2}. 
Taking account of the scaling factors, 
the average accretion rate of 2M\,1207 observed in May 2006 ($\dot{M} \sim 10^{-10}\,{\rm M_\odot/yr}$)   
yields an approximate value for the surface field, $B \approx 200$\,G. 
The same equation yields for a TTS with $0.8\,{\rm M_\odot}$, $1.5\,{\rm R_\odot}$, and
$\dot{M} \sim 10^{-8}\,{\rm M_\odot/yr}$ a surface field of $\approx 600$\,G. 
This validates the remark made by \cite{Scholz06.1} 
who, without citing numbers for the field strength, had argued that 
the magnetic field for a typical brown dwarf is expected to be roughly half that of a typical TTS. 
The magnetic field of TTS obtained this way is somewhat
lower than measured values, which are typically in the kilogauss range 
\citep[e.g.][ and references herein]{Yang05.1}. 
There are, however, considerable uncertainties connected with the use of the Koenigl-relation, 
such as the simplified assumption of a dipolar field and the location of the inner disk 
radius with respect to the corotation radius that determines the scaling factor. 
Given these ambiguities, 
the order of magnitude agreement between the estimated and observed fields is plausible. 

\subsection{Accretion versus activity in TWA brown dwarfs}\label{subsect:acc_act}

Given the presence of a magnetic field, inferred from the accretion rate, the absence of 
strong magnetic activity on 2M\,1207 is remarkable. A very stringent upper limit to its 
X-ray emission was placed by a $50$\,ksec {\em Chandra} observation \citep{Gizis04.1}. The 
radio emission was constrained to the lowest value among the brown dwarfs in the 
TWA \citep{Osten06.3}. In Table~\ref{tab:bds} we summarize the accretion and activity measures 
for all four TWA brown dwarfs. 

The data in Table~\ref{tab:bds} has been collected from the literature except for the X-ray
data for SSSPM\,1102. This object was identified as a probable substellar TWA member in a 
recent proper motion study of the Supercosmos Sky Survey data \citep{Scholz05.5}. SSSPM\,1102 is in
the field-of-view of an archived {\em XMM-Newton} observation (Obs-ID\,0112880201). We ran the
standard Science Analysis System source detection process, and find no X-ray source at the 
optical position of SSSPM\,1102. To estimate an upper limit to its X-ray luminosity, 
we measured the counts at the expected position of SSSPM\,1102
and in an annulus around this position (representing the background). After applying the appropriate area 
scaling factor and extracting the on-source time from the exposure map, 
an upper limit to the net source count rate was estimated using the algorithm of \cite{Kraft91.1}. 
The distance to SSSPM\,1102 has been given by \cite{Mamajek05.1} ($d \sim 43$\,pc). This yields
$L_{\rm x} < 5.3 \cdot 10^{26}$\,erg/s in the $0.5-10$\,keV band. 

It is constructive to compare 2M\,1207 to TWA-5\,B, the 
best-studied substellar object in the TWA after 2M\,1207. 
Matching their properties shows the dichotomy of one brown dwarf that accretes but is not 
magnetically active, and another brown dwarf that is magnetically active but does not accrete. 
Presumably, some mechanism suppresses activity in accreting substellar objects, as seems to be
the case for classical TTS \citep{Preibisch05.1,Telleschi06.1}. Possible explanations that have been
put forth in the context of classical TTS are changes in the magnetic field structure and/or heating process 
of the corona as a result of accretion. 
The absence of magnetic activity on 2M\,1207 is further surprising, as accreting substellar 
objects with similar magnetic field strength, 
e.g. V410\,Anon\,13, have been shown to be X-ray emitters 
\citep{Muzerolle00.2, Guedel06.1}.
For the substellar regime, the relation between accretion and X-ray luminosity has not yet been established 
on a statistical sample. If such a relation, 
as suggested by 2M\,1207 and TWA-5\,B, is confirmed, this would provide one more piece of evidence 
for the TTS-like character of brown dwarfs. 
X-ray and radio studies of the remaining two TWA brown dwarfs, both of which are non-accreting, should be useful.

\section{Conclusions}\label{sect:conclusions}

A substantial number of high-resolution spectroscopic observations has been collected for the brown 
dwarf 2M\,1207. These data have allowed us to study its accretion variability on
timescales from hours to $\sim 3$\,yrs.  
While small variations are obviously present throughout the rotation ($\sim 1$\,d), major 
changes in the structure of accretion may take place over months or years. 
A summary of all H$\alpha$ width measurements suggests a trend towards increasing accretion rate 
during the last three years. However, it is not clear if this trend reflects 
erratic changes or a systematic evolution. 
All in all, our dedicated monitoring has shown that 
the accretion pattern on 2M\,1207 is probably more complex than expected.
It comprises, longterm changes of the accretion rate and (maybe independently) of the flow structure
(months to years), variations on the timescale of the rotation rate (hours to days), and possibly
on even shorter timescales as suggested by our high temporal resolution data. 

2M\,1107 is among the lowest mass dwarfs with measured mass accretion rate. The values derived from
the various observations obtained in a time-span of $\sim 3$\,yrs, differ by nearly $2$\,dex, similar
to (and possibly explaining) the scatter observed in the relation between accretion rate and stellar mass. 
Whether accretion variability is decisive for shaping the $\dot{M}_{\rm acc} - M$ relation will be established in
future systematic spectroscopic studies across a wide mass range. 



\begin{table}
\begin{center}
\caption{Summary of H$\alpha$ width measurements for 2M\,1207.\label{tab:w10}}
\begin{tabular}{lrl} \hline
Date                  & $W_{\rm 10\,\%}$ & Reference \\ 
                      & [km/s]       &      \\ \hline
Apr 2002              & $\sim 300^\dagger$  & \protect\cite{Gizis02.2} \\
May 8, 2003           & $170...200$  & \protect\cite{Mohanty03.2} \\
Jan 29 - Feb 1, 2005  & $209...215$  & \protect\cite{Scholz05.4} \\ 
Mar 17 - 19, 2005     & $253...308$  & \protect\cite{Scholz05.4} \\
Mar 27 - 30, 2005     & $279...304$  & \protect\cite{Scholz05.4} \\
Feb 21 - 22, 2006     & $261...281$  & present work \\
Apr 11, 2006          & $253$        & present work \\
May 8 - 9, 2006       & $281...322$  & present work \\
May 16, 2006          & $320$        & \protect\cite{Whelan07.1} \\
\hline
\multicolumn{3}{c}{$^\dagger$ low-resolution data, no $10$\,\% width measured.} \\
\end{tabular}
\end{center}
\end{table}

\begin{table}
\begin{center}
\caption{Equivalent widths and FWHM for detected emission lines.\label{tab:eqw}}
\begin{tabular}{lrrr}\hline
Identification & $W_{\rm EQ}^{(a)}$              & typ.error$^{(a,b)}$         & FWHM$^{(c)}$ \\ 
               & \multicolumn{1}{c}{[\AA]} & \multicolumn{1}{c}{[\AA]}   & \multicolumn{1}{c}{[km/s]} \\ \hline
H$\beta$              & $4.5...47.1$  & $\sim 6$    & $180$ \\
H$\gamma$             & $0.3...13.5$  & $\sim 2$    & $180$ \\
H$\delta$             &    ...       & ...  & $168$ \\
H$\epsilon$+Ca\,II\,H &    ...       & ...  & $144$ \\
Ca\,II\,K             &    ...       & ...  & $18$ \\
He\,I\,$\lambda5876$  & $1.8 ... 9.1$ & $\sim 2$    & $30$ \\
H$\alpha$             & $145 ... 394$ & $\sim 45$   & $210$ \\
He\,I\,$\lambda6678$  & $0.9 ... 2.4$ & $\sim 2$    & $18$ \\
Ca\,II\,$\lambda8498$ & $0 ... 0.6$   & $\sim 0.35$ & $12$ \\
Ca\,II\,$\lambda8542$ & $0 ... 0.3$   & $\sim 0.25$ & $12$ \\
Ca\,II\,$\lambda8662$ & $0 ... 0.4$   & $\sim 0.3$  & $12$ \\ 
\hline
\multicolumn{4}{l}{$^{(a)}$ No equivalent widths are given for the higher Balmer lines} \\
\multicolumn{4}{l}{and for Ca\,II\,H+K because of the low continuum} \\
\multicolumn{4}{l}{$^{(b)}$ Errors for the equivalent width are estimated} \\ 
\multicolumn{4}{l}{ by assuming a $15$\,\% uncertainty in the} \\
\multicolumn{4}{l}{ continuum flux.} \\
\multicolumn{4}{l}{$^{(c)}$ measured in an average spectrum that} \\
\multicolumn{4}{l}{represents the typical line profile for 2M\,1207} \\
\multicolumn{4}{l}{during the observation.} \\
\end{tabular}
\end{center}
\end{table}

\begin{table}
\begin{center}
\caption{Activity and accretion measures of all known brown dwarfs in TWA.\label{tab:bds}}
\begin{tabular}{lrrrrrc} \hline
Designation & IR excess & $W_{\rm H\alpha;10\,\%}$ & $L_{\rm x}$    & $L_{\rm R}$ & FUV & Refs \\ 
            &           & [km/s]                   & [erg/s]        & [$10^{14}$\,erg/s/Hz]  &     & \\ \hline
TWA\,5B     &           & $162$                    & $4\,10^{27}$    & $< 2.0$  &         & (-,1,2,3,-) \\ 
2M\,1207    & $\surd$   & $170...320$              & $<1.2\,10^{26}$ & $< 0.98$ & $\surd$ & (4,5,6,3,7) \\
2M\,1139    & $-$       & $111$                    & ?               & ?        &         & (4,1,-,8,-) \\
SSSPM\,1102 &           & $194$                    & $<5.3\,10^{26}$ & $< 1.6$  &         & (-,1,9,3,-) \\
\hline
\multicolumn{7}{l}{(1) - \cite{Mohanty03.2}, (2) - \cite{Tsuboi03.1}, (3) - \cite{Osten06.3},} \\
\multicolumn{7}{l}{(4) - \cite{Riaz06.1}, (5) - see references in this paper, (6) - \cite{Gizis04.1},} \\
\multicolumn{7}{l}{(7) - \cite{Gizis05.1}, (8) - \cite{Burgasser05.1}, (9) - this paper.} \\
\end{tabular}
\end{center}
\end{table}


%
%

%
%

%
%

%
%

%
%

%
%

%
%

%
%

%
%



\acknowledgments

This paper is based on data collected under the ESO program 077.C-0323. 
BS acknowledges financial support from ASI contract ASI-INAF I/023/05/0.
BS wishes to thank S. Randich and S. Mohanty for useful discussions on the UVES 
pipeline analysis and brown dwarf emission line spectra. RJ and AS acknowledge 
support from an NSERC grant to RJ. 



{\it Facility:} \facility{VLT:Kueyen (UVES)}




\bibliographystyle{apj}
\bibliography{apj-jour,uves}




\clearpage

\end{document}